\documentclass[12pt,a4paper]{article}
\usepackage[compact]{titlesec}
\titlespacing{\section}{0pt}{*0}{*0}
\titlespacing{\subsection}{0pt}{*0}{*0}
\titlespacing{\subsubsection}{0pt}{*0}{*0}
\usepackage{array}
\usepackage{verbatim}
\usepackage{float}
\usepackage{bm}
\usepackage{amsmath}
\usepackage{graphicx}
\usepackage{setspace}
\usepackage{amssymb}
\usepackage{subfigure}
\usepackage[authoryear]{natbib}
\usepackage[left=2.5cm, right=2.5cm, bottom=1.6in,top=1.6in]{geometry}
\makeatletter

\providecommand{\tabularnewline}{\\}
\usepackage{bm}
\usepackage{amsmath,amsthm}
\usepackage{url}

\newtheorem{theorem}{Theorem}

\makeatother

\begin{document}

\title{Enhancing Bayesian risk prediction for epidemics using contact tracing}

\author{Chris P Jewell and Gareth O Roberts\\
Department of Statistics\\
University of Warwick\\
Coventry\\
CV4 7AL}
\maketitle
\begin{abstract}
Contact tracing data collected from disease outbreaks has received
relatively little attention in the epidemic modelling literature because it is thought to be unreliable: infection
sources might be wrongly attributed, or data might be missing due
to resource contraints in the questionnaire exercise. Nevertheless,
these data might provide a rich source of information on disease transmission
rate. This paper presents novel methodology for combining contact
tracing data with rate-based contact network data to improve
posterior precision, and therefore predictive accuracy. We present an advancement in 
Bayesian inference for epidemics that assimilates these data, and is robust to partial contact tracing.
Using a simulation study based on
the British poultry industry, we show how the presence of contact
tracing data improves posterior predictive accuracy, and can directly
inform a more effective control strategy.
\end{abstract}

\section{Introduction}

In a world in which people and animals move with increasing frequency
and distance, authorities must respond to disease outbreaks at maximum
efficiency according to economic, social, and political pressures.  Field 
epidemiologists are typically faced with making decisions based on imperfect and
heterogeneous data sources.  \citet{EEC92/119} Council Directive 92/119/EEC requires member states to identify
``other holdings...which may have become infected or contaminated''; 
this is achieved through contact tracing, performed by the competent authority -- 
for instance Defra's Framework Plan for Exotic Animal Diseases \citep{defraExoticDiseaseFramework}.  
Though resource limitations might prevent the implementation of efficient
case detection and removal based on contact tracing data (CTD), the presence
of CTD might provide a rich source of information,
both on contact frequency and probability of infection given a contact.
In a practical setting, however, rigorous collection of contact tracing
information from all detected cases is difficult to achieve; more
typically, it will be collected from as many individuals as resources
allow. Therefore, for a given epidemic the amount of contact tracing
data available for analysis is not guaranteed in advance, the challenge for the field
epidemiologist being how best to use these data to investigate disease dynamics, 
and inform disease control decisions.

This paper introduces a Bayesian approach to the assimilation of CTD into model-based
inference and control for infectious diseases based on spatiotemporal case data.  This can be applied to general classes of continuous time 
epidemic models, and we provide illustration using a flexible epidemic modelling framework as described in
\citet{OnRob99} and \citet{JewEtAl09c,JewEtAl2009b}.  Our methodology is particularly useful since it automatically adapts to
the information contained in CTD, over and above that contained in the adjunct timeseries.

In advance of an outbreak, network and spatiotemporal dynamical models of disease have become
a standard for evaluating the likely effect of various control measures,
and are now routinely used for contingency planning. Using demographic
data as well as case time series data, previous work has shown that
real time estimation of epidemic model parameters can potentially
be highly effective in informing decision-making \citep{CaucEtAl06,JewEtAl2009b}.
In the early stages of an epidemic case data are often sparse, though
this is when choosing the right course of control strategy is most
critical \citep{AnMay91}. Therefore, decisions must often be made
in the face of considerable uncertainty. For such a statistical decision
support tool, assimilating different sources of data is often helpful
in maximising statistical information, leading to more accurate model
predictions and improved decision-making throughout the epidemic \citep{PrEtAl10}.
Nevertheless, incorporating diverse data sources into a tractable
likelihood function commonly presents a methodological problem (see
for example \citet{DigEl95}).

Much has been written on how contact tracing may be used to decrease
the time between infection and detection (notification) during epidemics.
However, this focuses on the theoretical aspects of how contact tracing
efficiency is related to both epidemic dynamics and population structure.
It has been shown that, providing the efficiency of following up any
contacts to look for signs of disease is high, this is a highly effective
method of slowing the spread of an epidemic, and finally containing
it (see for example \citet{EamKeel03,KisGrKao05,KlFrHees06}). In
contrast, the use of CTD in inferring epidemic dynamics
does not appear to have been well exploited. During the UK outbreak
of foot and mouth disease in 2001, the Ministry of Agriculture, Farming,
and Fisheries (now Defra) inferred a spatial risk kernel by assuming
that the source of infection was correctly identified by the field
investigators, giving an empirical estimate of the probability of
infection as a function of distance \citep{FerDonAn01a,KeelEtAl01,Sav_et_al06}.
Strikingly, this shows a high degree of similarity to spatial kernel
estimates based on the statistical techniques of \citet{Dig06} and
\citet{Kyp07} without using contact tracing information. However,
\citet{CaucEtAl06} make the point that the analysis of imperfect
CTD requires more complex statistical approaches,
although they abandon contact tracing information altogether in their
analysis of the 2003 SARS epidemic in China. In human health, \citet{BlTr10}
devise a model that incorporates contact tracing for HIV-AIDS by dividing
cases into those detected by random sampling or contact tracing. This
allows them to estimate the proportion of undetected cases in the
population at a particular observation time, although the construction
of the model does not allow them to estimate the probability of infection
given a contact. 

Addressing this, we present a model
which uses contact tracing information between pairs of individuals
for time periods when it is available, and falls back on a Poisson
process likelihood for time periods when it is not. We illustrate
our methodology using an example from the UK agricultural industry.

In Section \ref{sec:Motivating-example}, we describe a motivating
example of a potential outbreak of avian influenza in the British
poultry industry. In Section \ref{sec:model-review}, the existing
approach to continuous time epidemic models is reviewed. Section \ref{sub:Incorporating-Contact-tracing}
then presents our new class of embedded model for incorporating CTD. In Section \ref{sub:MCMC-algorithm} we develop a reversible jump
Markov Chain Monte Carlo algorithm to estimate the posterior density,
showing the feasibility of exact Bayesian computation for problems
of this type. Section \ref{sec:Case-study} then applies
this methodology to the avian influenza example, showing how this
methodology might be used in the event of an outbreak of high pathogenicity
avian influenza.

\section{Motivating Example\label{sec:Motivating-example}}

One motivation for this paper stems from the example of a potential
avian influenza outbreak in the British poultry industry \citep{JewEtAl2009b}, using data from Defra's Great Britain Poultry Register as well as contact network data
\citep{Dent2008,ShEtAl08}. We identify three contact networks contained in the data with
different levels of information.  The company association network is ``static'': an edge is present if and only if two poultry holding belong to the same production company.  The feed mill delivery and slaughterhouse networks are ``dynamic'' meaning they contain edge frequency information -- for example, we know not only that two holdings are connected by a feed mill delivery, but how often a feed lorry runs between them.  Geographic coordinates for each holding allow us to consider non-network spatially related
transmission.  We also consider ``background'' infection
sources, accounting for infections unexplained by the other transmission
modes.

The presence
of explicit contact networks within the poultry industry presents
an interesting opportunity to investigate the possibility of including
CTD into the SIR-type epidemic model
\citep{KerMcK27}. Where contact frequency data is available,
we can view the networks as dynamic with contacts occurring according
to a Poisson process with intensity equal to the frequency along a
directed edge of the network. Normally this process is not observed,
and we only have a mean contact intensity between individuals to use as
covariate data in a statistical analysis. However, the collection
of CTD presents the possibility for making inference
from direct observations of the contact process, albeit for a select
period of time leading up to a case detection. 

In the typical livestock setting, CTD is gathered in response to the notification (ie case detection)
of an infectious premises (IP).  The resulting data are
a list of contacts (with time, source, destination, and type) that have been made in and out of the IP during
a period prior to the notification. The length of this
 period -- the ``contact tracing window'' -- during which
contact tracing is gathered is stipulated by policy, and is typically
longer than the expected infection to detection time for the disease
in question.

Though somewhat idealised here, these data provide us a platform 
from which to develop the methodology required to make use of CTD for inference on
epidemic models.  In livestock epidemics, CTD is not routinely passed to the modelling
community, and it is hoped that the results presented in this paper will encourage the authorities to do so.

\section{A review of continuous time epidemic models\label{sec:model-review}}

To begin, we provide an overview of the SINR epidemic model for heterogeneously-mixing
populations (see also \citet{JewEtAl09c}).
The SIR model is extended by considering the population to be composed
of individuals who, at any time $t$, exist in one
of four states: Susceptible, Infected, Notified, and Removed. Progression through 
these states is assumed to be serial: individuals begin as susceptible, become infected, are notified
(ie disease is detected), and are finally removed from the population
(either by death, or life-long immunity). The term {}``individual''
refers to an epidemiologically discrete unit, which might be a person
or animal, or might be of higher order, such as a household or farm
(as is the case for our HPAI example).  It is then assumed that individuals 
become infected via transmission modes $k=1,\dots,K$ which might be contact network or
spatial proximity to infected individuals.  Independent ``background'' sources of infection are also 
included to account for infection sources other than those explicitly modelled.

Conditioning on the initial
infective $\kappa$, individuals $j$ are assumed to become infected according
to a time-inhomogeneous Poisson process with instantaneous rate $\lambda_{j}(t)$
equal to the sum of the infection rates across all transmission modes from all infected and notified
individuals present in the population. Let $\mathcal{S}(t)$, $\mathcal{I}(t)$,
$\mathcal{N}(t)$, and $\mathcal{R}(t)$ be the sets of susceptible,
infected, notified, and removed individuals at time $t$, with the restriction that the entire population $\mathcal{P} = \mathcal{S}(t) \cup \mathcal{I}(t) \cup \mathcal{N}(t) \cup \mathcal{R}(t)$.  Furthermore, let $\bm{I}$,
$\bm{N}$, and $\bm{R}$ the corresponding vectors of individuals'
infection, notification, and removal times. Infection times are, of
course, never directly observed and we therefore treat $\bm{I}$ as
missing data. Since the aim of the methodology is to make inference
on an epidemic in progress observed at time $T_{obs}$, infected individuals are partitioned into
$j:N_j>T_{obs}$ occult (ie undetected) infections with right censored
notification and removal times, and $j:N_j \leq T_{obs}$ known (ie detected) infections.
Given the disease transmission parameters $\bm{\theta} = \left\{\epsilon, \bm{\beta}, \gamma, \bm{p} \right\}$ (see below), the likelihood function (here denoted by $L_A(\cdot)$) for the epidemic process up to the analysis
time $T_{obs}$ is 
\begin{eqnarray}
L_{A}(\bm{I}, \bm{N},\bm{R}|\bm{\theta}) & = & \prod_{j=1,j\ne\kappa}\left(\lambda_{j}(I_{j}^{-})\right)\cdot\exp\left[-\int_{I_{\kappa}}^{T_{obs}}\left(\sum_{j \in \mathcal{P} ,j\ne\kappa}\lambda_{j}(t)\right)dt\right]\nonumber \\
 & \times &  \prod_{j:N\leq T_{obs}}f_{D}\left(N_{j}-I_{j}\right) \times \prod_{j:N_j>T_{obs}}\left[1-F_{D}\left(T_{obs}-I_{j}\right)\right] \label{eq:generic-likelihood}
 \end{eqnarray}

where $f_{D}(\cdot)$ is the pdf, and $F_{D}(\cdot)$ the corresponding
cdf, for the infection to notification time. $\lambda_{j}(t)$ is further
decomposed
\begin{equation}\label{eq:lambdaj}
\lambda_{j}(t)= \epsilon + \sum_{k=1}^{K} \left[ \sum_{i\in\mathcal{I}(t)}\lambda_{ij}^{(k)}(t) + \gamma \sum_{i\in\mathcal{N}(t)} \lambda_{ij}^{(k)}(t) \right]
\end{equation}
where $\epsilon$ represents background infection rate common to all individuals, and $\gamma$ measures the effect of quarantine measures imposed upon notified individuals.  A wide range of transmission modes can then be specified, for example
\begin{equation}\label{eq:txModes}
\lambda_{ij}^{(k)}(t) =  \begin{cases}q(i;\bm{\xi},t)s(j;\bm{\eta})\Psi_k(i,j;\psi) & \mbox{for spatial proximity} \\
							   q(i;\bm{\xi},t)s(j;\bm{\eta})\beta_kc_{ij}^{(k)} & \mbox{for \emph{associative} networks} \\ 
							   q(i;\bm{\xi},t)s(j;\bm{\eta})p_kr_{ij}^{(k)} & \mbox{for \emph{frequency} networks}. \end{cases}
\end{equation}

In these examples, the function $0 \leq q(i;\bm{\xi},t) \leq 1$ represents the infectivity of individual
$i$ at time $t$, and $0 \leq s(j;\bm{\eta}) \leq 1$ the susceptibility of individual $j$ which is assumed constant.
The rate of spatial (or environmental) transmission is given by $\Psi_k(i,j;\psi)$ and would typically be a function of Euclidean distance between $i$ and $j$.
Network rates of transmission are represented by $\beta_kc_{ij}^{(k)}$
and $p_kr_{ij}^{(k)}$. The important distinction between
the latter two terms is that $c_{ij}^{(k)}$
represents network \emph{associations} (0 or 1) between $i$ and $j$
with associated infection rate $\beta_k$, whereas $r_{ij}^{(k)}$
represents a vector of (potentially infectious) network \emph{contact rates} with associated
probability $p_k$ that a contact results in an infection.

We remark that, if contacts are assumed to occur according to some underlying Poisson process
with rate $r_{ij}^{(k)}$, infections occur according to a thinned
version of this Poisson process, the thinning being governed by $p_k$.
We assume that the contact rate is \emph{a
priori }independent of the probability of the infection being transmitted. Thus, a contact between infected
individual $i$ and susceptible $j$ will not necessarily result in
$j$ being infected, just that there is some positive probability of infection
being transmitted.  Considering a putative full model containing several networks, the information contained in $\bm{p}$
therefore gives a direct measure of risk for each contact network in
$\bm{r}_{ij}$.

\section{Incorporating Contact tracing data\label{sub:Incorporating-Contact-tracing}}

Contact tracing data, $\bm{\mathcal{C}}$, represents a list of all known contacts, both
incoming and outgoing, that have occurred between newly detected cases $j$
and all other individuals during a contact tracing window defined as the interval $[T_j^c,N_j)$.   In other words, these data provide a means to observe the underlying contact process driving the infectious process as discussed in Section \ref{sec:model-review}.  

Let $C_{ijk}(t)$ be a right continuous counting process describing the number of contacts that have occurred between infected individuals $i$ and $j$ along network $k$ up to time $t$, with $\Delta C_{ijk}(t) = \lim_{\delta\downarrow 0} \left[C_{ijk}(t) - C_{ijk}(t-\delta)\right]$.  
If it were possible to obtain CTD for all time, as well as perfectly observing individuals' infection states, then inference on $\bm{p}$ could be made using a geometric model with likelihood denoted by $L_\Omega(\cdot)$. For example

\begin{eqnarray}\label{eq:geometric-model}
L_\Omega(\bm{I}, \bm{N}, \bm{R}, \bm{\mathcal{C}} | \bm{p}) & = & \prod_{j\in\mathcal{P}} \exp \left[ \int_{I_\kappa}^{I_j \wedge T_{obs}} \sum_{i\in\mathcal{Y}(t)} \sum_{k=1}^{K} Z_{ijk}(t)dC_{ijk}(t) \right] \nonumber \\
 & \times &  \prod_{j:N\leq T_{obs}}f_{D}\left(N_{j}-I_{j}\right) \times \prod_{j:N_j>T_{obs}}\left[1-F_{D}\left(T_{obs}-I_{j}\right)\right]
\end{eqnarray}
with \[ Z_{ijk}(t) =  \bm{1}[I_j = t] \log(q(i;\bm{\xi},I_j-I_i)s(j;\bm{\eta})p_k) + (1-\bm{1}[I_j \geq t])\log(1-q(i;\bm{\xi},I_j-I_i)s(j;\bm{\eta})p_k)  \] and $\mathcal{Y}(t) = \mathcal{I}(t) \cup \mathcal{N}(t)$.

The limitation to this approach is, of course, that CTD is only available for the contact tracing window and so the naive application of such a likelihood will
give a biased estimate of $\bm{p}$.  

To address this, we divide the epidemic into periods of time for which CTD is observed, and periods for which it is not.  For each individual $j$ in the population, let $\mathcal{W}_j$ be the set of times for which contact tracing is observed (ie the contact tracing window), and $\mathcal{W}_j^c$ the set of times for which it is not.  Then, the $\sigma$-algebras $\mathcal{F} = \sigma(\Delta C_{ijk}(t); t\in\mathcal{W}_j)$ and $\mathcal{G} = \sigma(\Delta C_{ijk}(t); t\in\mathcal{W}_j^c)$ represent the information contained in CTD, and that lost by not having CTD, respectively. 

\begin{theorem}\label{th:main}
Where CTD is only available for contact tracing windows $\mathcal{W}_j$, if the underlying contact rates $r_{ij}^{(k)}, k=1,\dots,K$ are known, the likelihood with respect to $\mathcal{F}$ is obtained by taking an expectation with respect to $\mathcal{G}$ such that
\begin{eqnarray}\label{eq:ctLikelihood}
L_{\mathcal{F}}(\bm{I},\bm{N},\bm{R},\bm{\mathcal{C}} | \bm{\theta}) & = & \prod_{j:I_j \in \mathcal{W}_j} \exp \left[ \int_{\mathcal{W}_j} \sum_{i\in \mathcal{Y}(t)} \sum_{k=1}^{K} Z_{ijk}(t) dC_{ijk}(t) \right]\nonumber \\
& \times & \prod_{\substack{j:I_j\in \mathcal{W}_j^c \\ j\ne\kappa}} \sum_{i\in\mathcal{Y}(I_{j}^{-})}\lambda_j(I_j^-)\cdot\exp\left[-\int_{\mathcal{W}_j^c}\lambda_j(t)dt\right] \nonumber\\
 & \times &  \prod_{j:N\leq T_{obs}}f_{D}\left(N_{j}-I_{j}\right) \times \prod_{j:N_j>T_{obs}}\left[1-F_{D}\left(T_{obs}-I_{j}\right)\right].
\end{eqnarray}

See Supplementary Material A for proof.
\end{theorem}

Theorem \ref{th:main} relies on knowing the underlying contact rate for a particular network.  For associative and spatial transmission modes (and indeed background pressure) the contact rate between individuals, $r_{ij}^{(k)}$, is not known and these cases do not therefore fit into our contact tracing framework.  However, since the overall infection rate $\beta_k = p_k r_{ij}^{(k)}$, where it is assumed that $r_{ij}^{(k)}$ is constant with respect to both $i$ and $j$, we can regard infections via these modes as occurring due to an unobserved contact process (Equation \ref{eq:txModes}).  The likelihood function in Equation \ref{eq:ctLikelihood} is therefore sufficiently flexible to include these transmission modes in the $\mathcal{W}_j^c$ related terms.  

Finally, we highlight that the likelihood function is written in terms of the data $\bm{I}$, $\bm{N}$, and $\bm{R}$, given the transmission parameters $\bm{\theta}$.  However, the epidemic process means that infection times are never directly observed.  We therefore treat $\bm{I}$ as missing data, using Bayesian data augmentation techniques described in the next section.

\section{Bayesian Inference\label{sub:MCMC-algorithm}}

A Bayesian approach permits the incorporation of prior information early in the epidemic, 
and provides
a natural way to deal with $\bm{I}$, the vector of censored infection times.

Independent priors are assigned to the model parameters
$\bm{\theta}=\left\{ \epsilon,\bm{p},\bm{\beta},\gamma, \bm{\psi},\bm{\eta},\bm{\xi}\right\} $.
Typically, Gamma distributions are used for rate parameters (ie $\epsilon$,
$\bm{\beta}$, $\gamma$, $\phi$) with Beta distributions used for probability
parameters (ie $\bm{p}$), as demonstrated in
Section \ref{sec:Case-study}.

The main features
of the adaptive
reversible jump MCMC algorithm used to fit the model are outlined here, with full implementational
details available in the Supplementary Material B. An adaptive multisite update
\citep{HaarSakTam01} is used for the disease transmission parameters $\bm{\theta}$, and avoids the necessity for time consuming pilot tuning runs. This is useful to speed up the
algorithm usage in a real time setting. The implementation of reversible
jump \citep{Green95} is two fold. Firstly, the dimension of $\bm{I}$
is allowed to expand or contract according to either and addition
or deletion update step for occult infections from the parameter space
\citep{JewEtAl09c}. This allows the algorithm to explore the possibility
occult infections present at the time of analysis consistent with
the model.  Secondly, infection events (including occults) are attributed to either a traced
contact related to the $\mathcal{W}_j$ components of transmission, or spatial/untraced
infections related to the $\mathcal{W}^c_j$ components. The switching in and out of the sets $\mathcal{W}_j$  comprises a reversible
jump move, since it results in the infection appearing in different
parts of the likelihood. The update step for an infection time therefore proposes the reversible
jump with probability 0.5, and a contact to
contact, or non-contact to non-contact move otherwise.

Shared memory parallel computing (GNU C++ using OpenMP) is used to calculate the likelihood, ensuring that the algorithm runs in an overnight timeframe \citep{JewEtAl09c}.

\section{Case study\label{sec:Case-study}}

To illustrate how this contact tracing methodology might
be used in practice, we return to the dataset presented in
Section \ref{sec:Motivating-example}. First we
describe the special case of HPAI within the
British poultry industry. Since no epidemic outbreak of HPAI has yet
occurred in Britain, we provide two simulation studies to demonstrate
our approach for the given model. We then
demonstrate how posterior information changes in response to the
amount of data available from an epidemic, both in terms of the length
of the timeseries and the presence or absence of CTD.
Finally we present a simulation study of
4 outbreak scenarios in which different surveillance strategies are
employed for early detection of new infections.

\paragraph{The model\label{sub:The-model}}

For this example, four modes of contact (Feedmill,
Slaughterhouse, and Company Networks, and Spatial (environmental)
transmission) contribute to the infection rate between infected or
notified individual (ie farm) $i$ and susceptible $j$. For the Feedmill
and Slaughterhouse networks, contact frequency information is available, whereas for the
Company network only the presence or absence of a business association
is known. The spatial transmission rate is then parameterised
as a function of the Euclidean distance between the two individuals, centred at 5km to improve MCMC mixing and aid parameter
interpretability. The transmission modes are therefore
\begin{eqnarray}
\lambda_{ij}^{(1)}(t) & = & q(i;\bm{\xi},t)s(j;\bm{\eta})p_1r_{ij}^{(FM)}\bm{1}[i \in \mathcal{I}(t)] \nonumber \\
\lambda_{ij}^{(2)}(t) & = & q(i;\bm{\xi},t)s(j;\bm{\eta})p_2r_{ij}^{(SH)}\bm{1}[i \in \mathcal{I}(t)] \nonumber \\
\lambda_{ij}^{(3)}(t) & = & q(i;\bm{\xi},t)s(j;\bm{\eta})\beta_1c_{ij}^{(CP)}\bm{1}[i \in \mathcal{I}(t)] \nonumber \\
\lambda_{ij}^{(4)}(t) & = & q(i;\bm{\xi},t)s(j;\bm{\eta})\beta_2e^{-\psi(\rho_{ij} - 5)} \nonumber
\end{eqnarray}
where $r_{ij}^{FM}$ and $r_{ij}^{SH}$ are contact \emph{frequencies}
for Feedmill and Slaughterhouse contacts respectively with associated
probabilities of infection $p_{1}$ and $p_{2}$, $c_{ij}^{CP}$ is
0 or 1 depending on whether a business association exists between
$i$ and $j$ with associated rate of infection $\beta_{1}$, $\rho_{ij}$
represents the Euclidean distance in kilometres between $i$ and $j$ with an exponential
distance kernel with decay $\psi$ and parameter $\beta_{2}$ interpreted as the rate of infection between individuals 5km apart.  The indicator function $\bm{1}[i \in \mathcal{I}(t)]$ removes the network transmission modes during the notified period, reflecting statutory movement restrictions on the infected farm \citep{EEC92/119}.

The infectivity function $q(i;\bm{\xi},t)$ is defined 
\[
q(i;\bm{\xi},t)=\begin{cases}
\frac{e^{\nu t}}{\mu+e^{\nu t}} & t \geq 0, I_i < t < N_i \\
1 & t \geq 0, N_i < t < R_i \\
0 & \mbox{otherwise}\end{cases}
\]
with $\nu=1.3$ and $\mu=60$ is assumed, determined by fitting $q(i;\bm{\xi},t)$ to expert opinion.
$\bm{\eta}$ is a 10-dimensional vector of susceptibilities for each
production type, such that $s(j;\bm{\eta})$ returns the susceptibility of the major production species on farm $j$.
In our example, we assume that broiler chickens are the most susceptible, and set this to 1.  All other elements of $\bm{\eta}$ are therefore
susceptibilities relative to broilers.
As in our previous work, $f_D(\cdot)$ is given by 
$
f_D(t)=ab\cdot\exp\left[bt-a\left(e^{bt}-1\right)\right]
$ 
with $a=0.015$ and $b=0.8$ assumed, giving a mean infectious period of 6 days as determined from expert opinion.  Prior distributions for the remaining parameters were chosen as suggested in Section \ref{sub:MCMC-algorithm},
and are shown in Table \ref{tab:CT.simParms}.

\paragraph{Posterior information\label{sub:Posterior-information}}

To illustrate how the acquisition of CTD
can improve posterior precision, we use a simulated epidemic on the
GBPR dataset as described in Section \ref{sec:Motivating-example},
using the model described in Section \ref{sec:model-review}. The
epidemic is realised using a stochastic simulation based on the Doob-Gillespie
algorithm \citep{Gil76} applied at the level of individual contacts
(to generate CTD), with the extension that retrospective
sampling is used to determine whether a contact
results in an infection \citep{JewEtAl2009b}.

Here we investigate how the assimilation of improves inference by increasing posterior
precision. We base our test analysis on a typical simulated epidemic lasting 109 days in which
350 farms become infected, using the parameter values
presented in Table \ref{tab:CT.simParms} (see also Supplementary Material C, Figure 1). Two timepoints were analysed,
day 40 (incomplete epidemic) and day 109 (complete epidemic), as shown
in Table \ref{tab:parmEpi}. CTD for a 21 day period
preceeding a notification event was recorded for each notified individual.
The algorithm was then run for the two timepoints during the
epidemic, with and without the associated CTD.

\begin{figure}[H]
\centering
\subfigure[Day 40]{\includegraphics[width=1.0\textwidth]{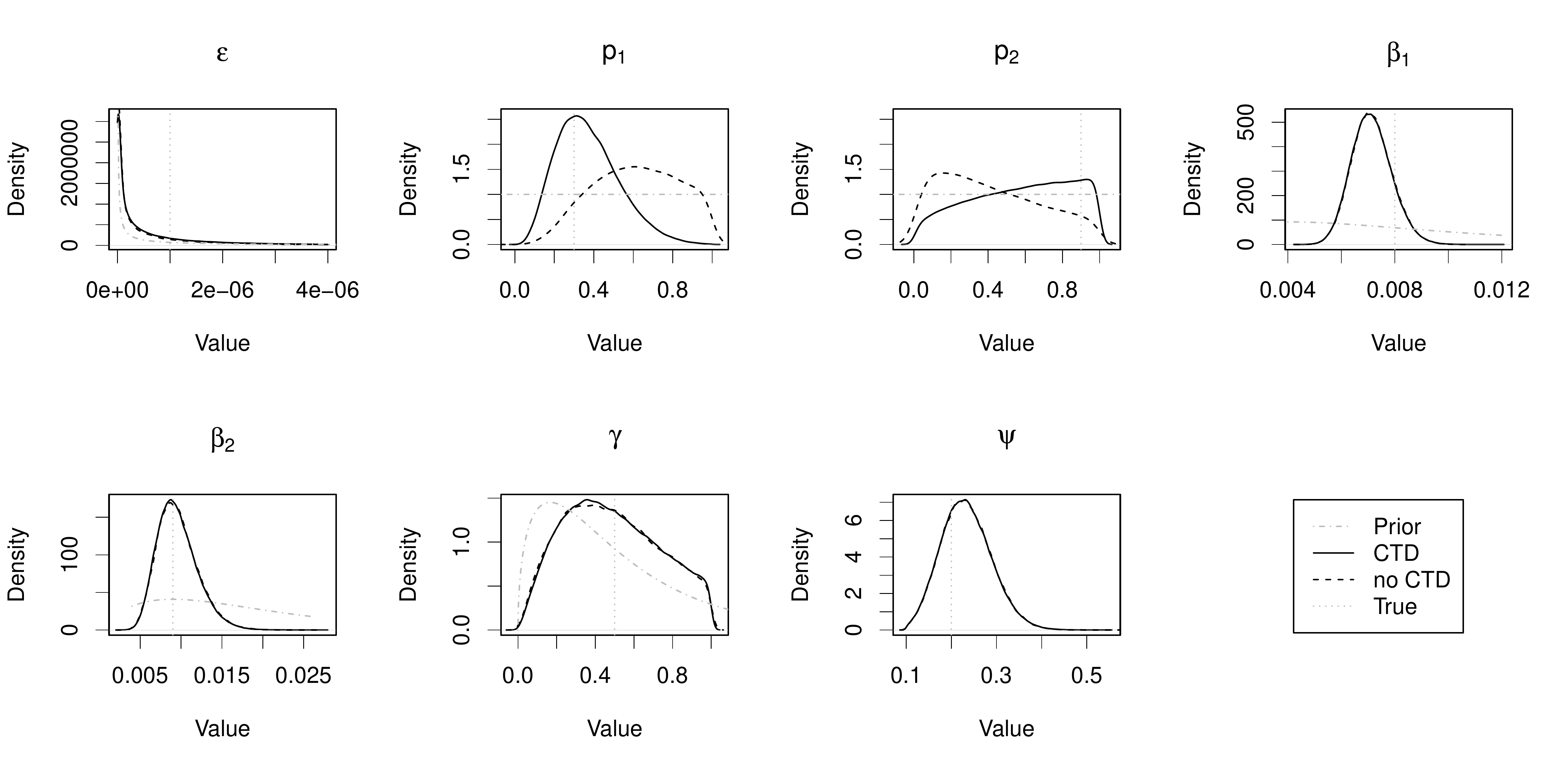}}
\subfigure[Day 109]{\includegraphics[width=1.0\textwidth]{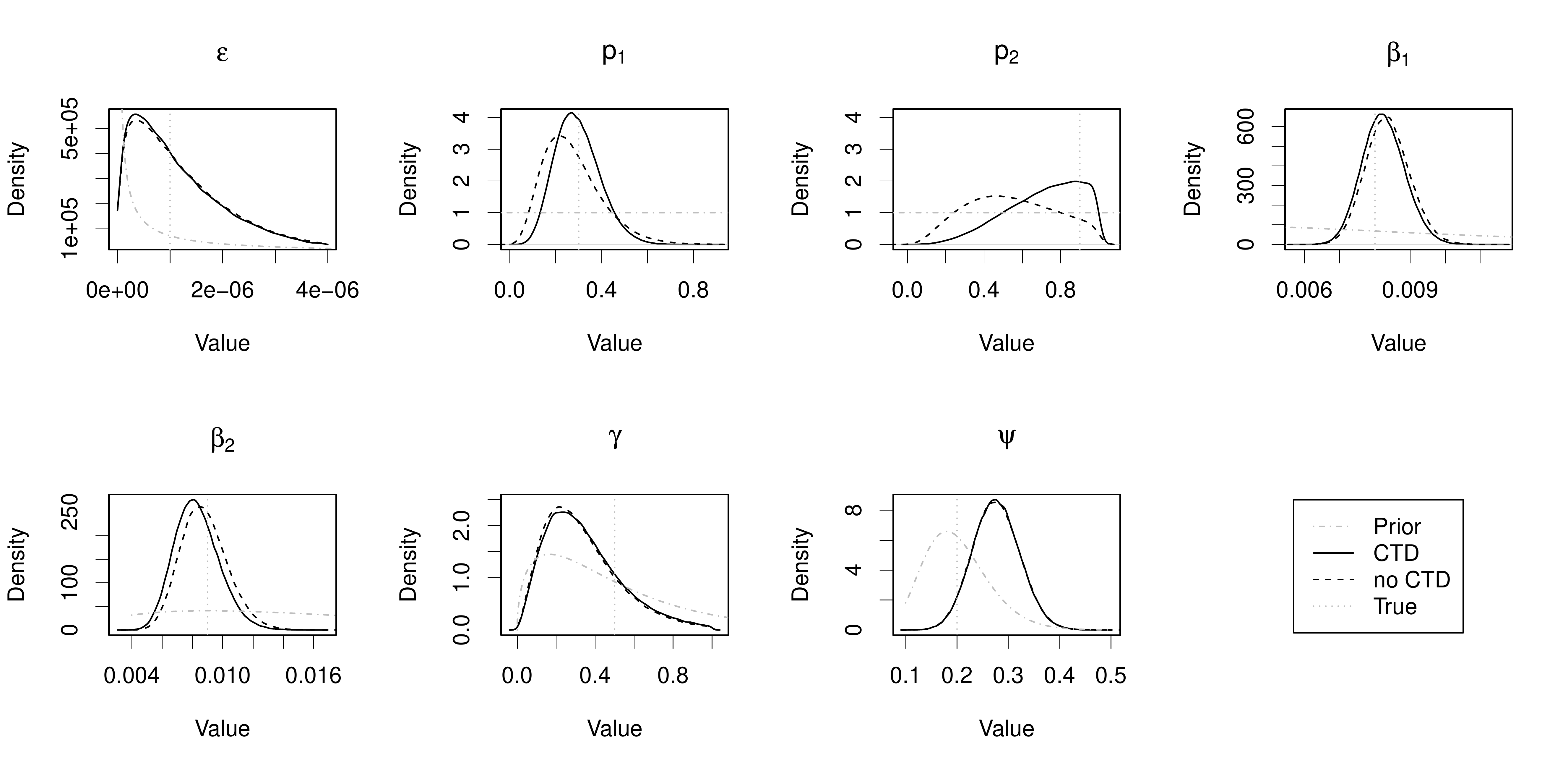}}
\caption{\label{fig:densities-trunc}Kernel density estimates of the marginal
posterior distributions of $\epsilon$, $p_{1}$, $p_{2}$, $\beta_{1}$,
$\beta_{2}$, $\gamma$, $\psi$ for Day 40 and Day 109.}

\end{figure}

The density plots for parameters $\epsilon$, $p_{1}$, $p_{2}$,
$\beta_{1}$, $\beta_{2}$, $\gamma$, and $\psi$ are shown in
Figure \ref{fig:densities-trunc}. Of particular
interest are the plots for the probability parameters $p_{1}$ and
$p_{2}$, which are, of course, directly informed by the contact tracing
information. It is immediately apparent that the addition of contact
tracing data affects the marginal posteriors of these parameters,
with an increase in precision in each case. The other parameters are
far less affected, with only minor differences apparent in the complete
epidemic analysis. Density plots for the components of $\bm{\eta}$ and histograms of the posterior number
of occults present on day 40 are shown in the Supplementary Material C, Figures 2-4.

\paragraph{Practical application\label{sub:Practical-application}}

To test a prospective application of our methodology, we
focus on the statistical detection of occult infections. We postulate that this provides
a method to target limited surveillance resource to the
most likely infected individuals.
A simulation study was performed in which 4 surveillance
strategies were tested. In the ``Reactive'' strategy, no active surveillance is performed,
and cases are notionally detected and reported by the farmer. The 3 remaining strategies
use active (pre-emptive) surveillance to look for disease as well
as case detection by the farmer, with the strategy being implemented
at day 14, mimicking the application of such a strategy once it has
become apparent that an epidemic is taking off. ``Random'' surveillance
represents a strategy in which a random sample of size $z$ holdings
within the statutory 10km surveillance zone are visited and tested daily \citep{defraExoticDiseaseFramework}. The ``Bayes'' targeted strategy uses our algorithm (without contact
tracing data) on a daily basis to rank holdings in terms of occult
probability; the top $z$ holdings are then visited and tested for
disease. The ``Bayes-CT'' strategy then adds in CTD to
target surveillance similarly. For the purposes of our simulation
study, we assume a surveillance resource of $z=15$. We also assume
that each surveillance team has at their disposal a perfect test for
the disease - ie 100\% sensitivity and specificity - and that if positive,
the farm is culled immediately. For each strategy, 500 epidemic realisations
with random index cases were generated to integrate
over the stochasticity in the epidemic process and study the efficacy
of each surveillance strategy. These were further conditioned to involve
at least 2 individuals and last greater than 14 days, such that the epidemic could be said to have
``taken off''. A flow diagram of the simulation study is shown
in the Supplementary Material C, Figure 5.

\begin{figure}[H]
\centering
\subfigure[The distribution of the
logarithm of total number of culled premises (reactive culling plus
active surveillance culls), log(mean number culled) shown by dashed lines]{\includegraphics[width=0.6\textwidth]{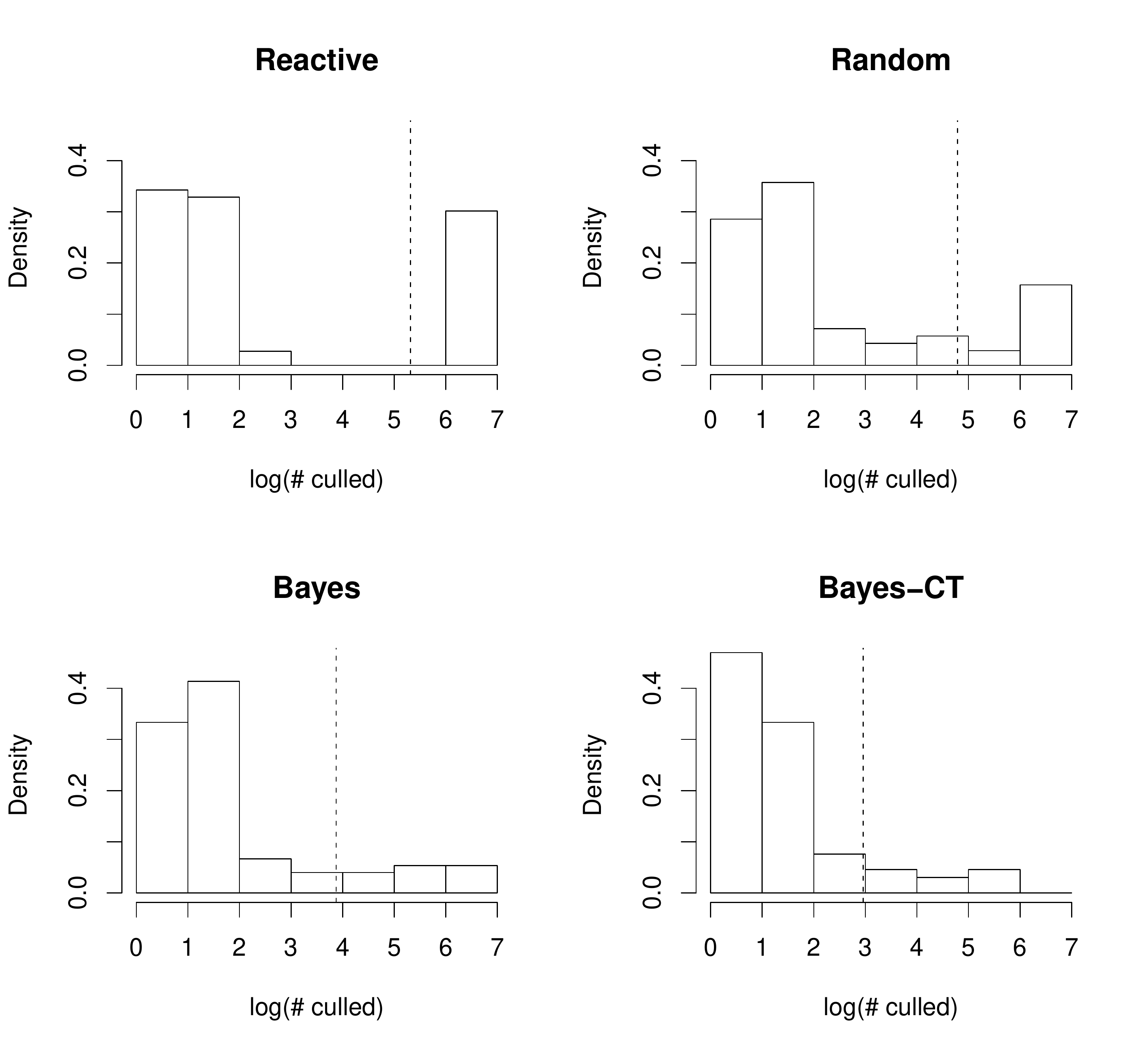} \label{fig:epiSizes}}
\subfigure[The distribution of epidemic duration, mean duration shown by dashed lines.]{\includegraphics[width=0.6\textwidth]{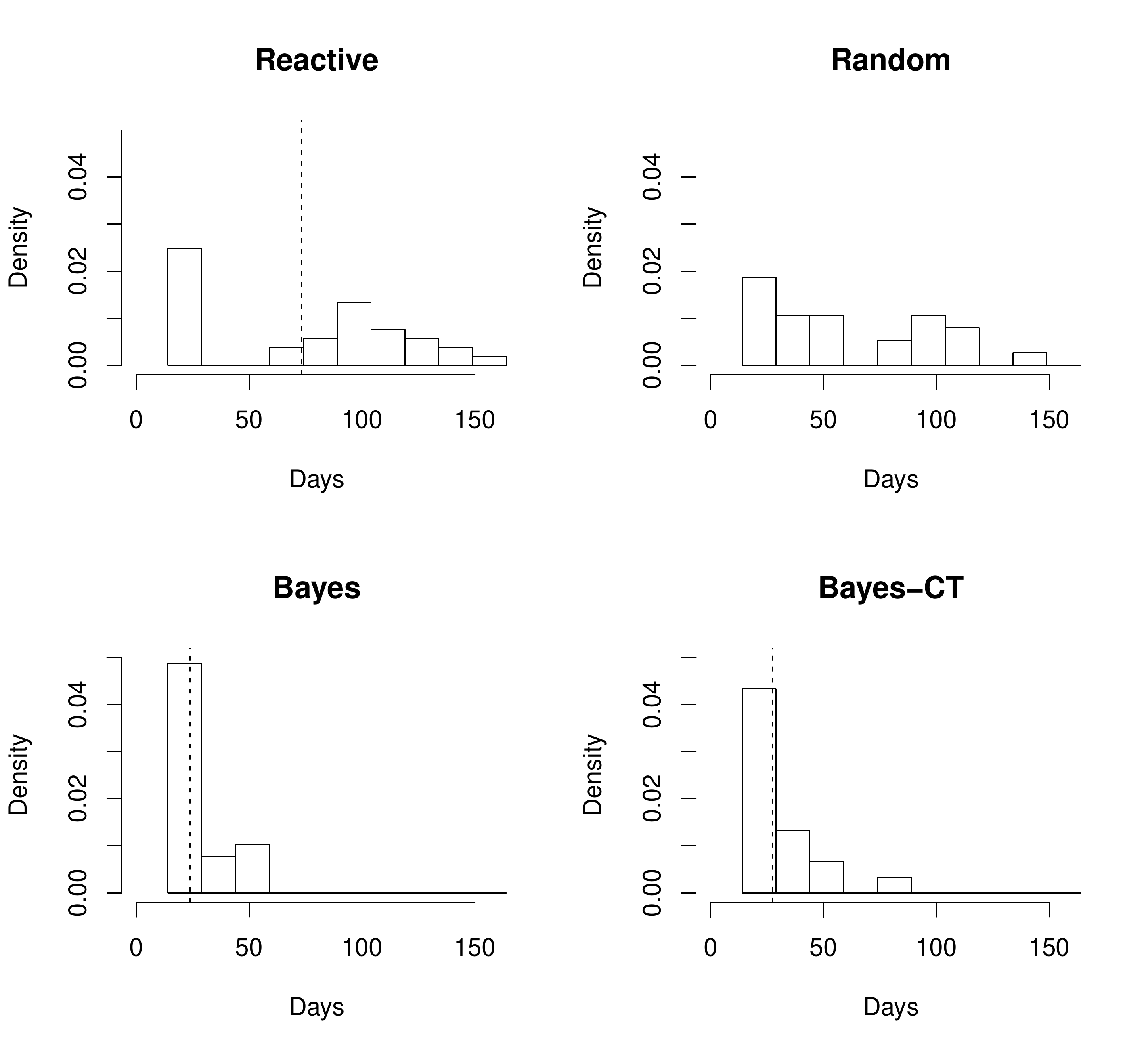} \label{fig:epiDuration}}
\caption{Histograms of total number of farms culled and epidemic duration under the 4 control strategies in the simulation study.\label{fig:epi-size-distribution}}

\end{figure}

As metrics for comparing the performance of the 4 surveillance strategies,
the number of culled holdings (both as a result of Reactive and surveillance
detection) and epidemic duration (ie time from the first infection
to the last cull with no further infected or notified holdings) were
used. These results are summarised in Table \ref{tab:epi-size-distribution} 
and Figure \ref{fig:epi-size-distribution}, and indicate
that active surveillance as implemented here is effective in reducing
the probability of large scale outbreaks.  The Random strategy, whilst successful in reducing
the mean size of the outbreak, does not appreciably affect the probability of a large epidemic.
Using Bayes targetting effects both
of dramatically reducing the mean epidemic size \emph{and} the probability of a large epidemic.
The addition of CTD improves the efficacy of surveillance further:
the Bayes strategy reduces the mean epidemic size by 4-fold as compared
to the Reactive strategy, the Bayes-CT strategy reduces it by 10-fold. A similar picture is reflected
in the results for epidemic duration, with the presence of a surveillance
strategy greatly reducing the mean. As before, the effect of Bayes
targeting is profound, with the presence of CTD reducing
the mean variance of the duration distribution. These results suggest that
 Bayes targeting of surveillance using case incidence
data \emph{and} contact tracing may have much to offer in disease
control resource prioritisation.

\section{Discussion}

The results presented in the previous section show contact tracing data is a
useful addition for inference and prediction on SIR type epidemic models. The purpose of
this methodological innovation is demonstrated
well in Figure \ref{fig:densities-trunc} 
showing an increase in posterior precision in response to
the acquisition of the contact tracing information in parameters $p_{1}$
and $p_{2}$. Of particular interest is the behaviour of the posterior
in response to differing amounts of data, and for this reason the
density plots should be considered in conjunction with Table \ref{tab:parmEpi}.
For day 40, only 1 out of 4 slaughterhouse contacts resulted
in infection, compared to 3 out of 28 for feedmill contacts. Importantly, the low contact
frequency of the slaughterhouse network (see Supplementary Material C, Figure 6; Supplementary Material D) means that the marginal posterior
for $p_{2}$ is poorly informed without the addition of contact tracing
data. Conversely, the full epidemic dataset on day 109 contains more
contact observations, reflected in the narrower posterior distributions. Here, there is perhaps little
difference in the marginal posteriors for $p_{1}$ with and without
CTD, though the effect on $p_{2}$ remains marked.

Despite its simplistic setup, the results of the surveillance strategy
simulation study are encouraging, lending evidence to Bayes guided
surveillance being highly effective for reducing
infection to detection time, and hence abrogating the spread of an
epidemic. Importantly, this methodology provides a way of immediately
linking targeted surveillance to changes in outbreak dynamics due to
unexpected behaviour of a new strain of the disease, changes in control
strategy, or changes in underlying population behaviour -- changes
that may well not be apparent to the naked eye in an evolving epidemic
timeseries. 

Given the marked influence that CTD has on the posterior
parameter estimates, care must be taken to avoid biasing the results
through the use of biased datasets. The structure of the likelihood assumes
independence between the data contained in individual contact tracing
questionnaires (ie between individuals), and therefore the model is
robust to absence of contact tracing from random cases. This is important as the analysis is unbiased in the case that the
speed of the disease process exceeds the capacity to collect contact
tracing data. Source of bias, however,
may well arise through a case's reluctance to declare certain contacts, and therefore carefully
designing tactful contact tracing questionnaires should be of high priority.
Nevertheless, in our Bayesian setup, the model extends naturally to
include expert opinion on the level of such reporting bias, so an
attempt to correct for it can be made.

In a UK livestock setting, although CTD is collected
and used informally by field operatives to identify at risk farms,
there currently appears to be no formal dissemination
of this data to analysts. However, given the strong evidence for its
use presented here, we strongly recommend that it should be made routinely
available together with case data. Given reliable sources of such
data, therefore, we envisage that real time inference and risk prediction
for epidemics will become commonplace in reactive disease control
strategy.

\textbf{Supplementary Materials}: Supplementary
Material contains the proof of Theorem \ref{th:main}, technicalities of the rjMCMC algorithm used in Section \ref{sub:MCMC-algorithm},
supplementary figures, and an intuition for the information gained using CTD.

\textbf{Acknowledgements}{: We thank Professors Laura Green and Matt Keeling for helpful discussions, BBSRC for funding for this work. We thank Defra for supplying demographic data, and also for their valuable discussions on the nature of contact tracing data.}

\bibliographystyle{apalike}
\bibliography{general}

\section*{Tables and Figures}

\begin{table}[H]
\noindent \begin{centering}
\begin{tabular}{ccc}
\hline 
\textbf{Parameter} & \textbf{True Value} & \textbf{Prior}\tabularnewline
\hline 
$\epsilon$ & $1^{-6}\mbox{day}^{-1}$ & Gamma$(0.15,5000)$\tabularnewline
$p_{1}$ & 0.3 & Beta$(1,1)$\tabularnewline
$p_{2}$ & 0.9 & Beta$(1,1)$\tabularnewline
$\beta_{1}$ & $0.008\mbox{day}^{-1}$ & Gamma$(2.048,256)$\tabularnewline
$\beta_{2}$ & $0.009\mbox{day}^{-1}$ & Gamma$(2,111)$\tabularnewline
$\gamma$ & $0.5$ & Gamma$(1.5,3)$\tabularnewline
$\psi$ & $0.2\mbox{km}^{-1}$ & Gamma$(10,50)$\tabularnewline
$\eta_{1}$ & 1 & fixed\tabularnewline
$\eta_{2}$ & 0.6 & Gamma$(1,10)$\tabularnewline
$\eta_{3\dots10}$ & 0.3 & Gamma$(1,10)$\tabularnewline
\hline
\end{tabular}
\par\end{centering}

\caption{\label{tab:CT.simParms}Transmission parameter {}``true'' values
used to simulate the test epidemic for contact tracing.}

\end{table}

\begin{table}[H]
\begin{centering}
\begin{tabular}{>{\centering}p{0.75in}>{\centering}p{0.75in}>{\centering}p{0.75in}c>{\centering}p{0.75in}>{\centering}p{0.75in}>{\centering}p{0.75in}}
\hline 
\textbf{Time / days}  & \textbf{Notified infections}  & \textbf{Occult infections}  & \multicolumn{2}{>{\centering}p{1.5in}}{\textbf{Contacts resulting in infection}} & \multicolumn{2}{>{\centering}p{1.5in}}{\textbf{Contacts not resulting in infection}}\tabularnewline
\hline
 &  &  & \multicolumn{1}{>{\centering}p{0.75in}}{Feedmill} & \multicolumn{1}{c}{S'house} & \multicolumn{1}{c}{Feedmill} & \multicolumn{1}{c}{S'house}\tabularnewline
\cline{4-7} 
40 & 159 & 19 & 3 & 1 & 25 & 3\tabularnewline
109 & 350 & 0  & 7 & 5 & 78 & 12\tabularnewline
\hline
\end{tabular}
\par\end{centering}

\caption{\label{tab:parmEpi}The state of the epidemic at each observation
time. True number of \emph{epidemiologically relevant }(ie originating
at infectious individuals) contacts are known from the simulation
algorithm.}

\end{table}
\begin{table}[H]
\noindent \begin{centering}
\begin{tabular}{ccc}
\hline 
\textbf{Strategy} & \textbf{Mean \# culled (95\% CI)} &  \textbf{Mean duration (95\% CI)}\tabularnewline
\hline
SOS & 203.5 (2,727) & 73.2 (14.5, 147.2)\tabularnewline
Random & 120.1 (2,709) & 60.0 (16.8, 122.4)\tabularnewline
Bayes & 48.1 (2,521) & 23.8 (14.3, 50.5)\tabularnewline
Bayes-CT & 19.3 (2,204) & 27.3 (14.2, 64.8)\tabularnewline
\hline
\end{tabular}
\par\end{centering}

\caption{\label{tab:epi-size-distribution}Mean number of holdings culled and mean epidemic duration (time from first infection to last removal) for
each surveillance strategy, conditional on the epidemic lasting longer than 14 days.  CI = credible interval.}

\end{table}

\end{document}

\appendix

\title{Contact tracing enhanced Bayesian risk prediction for epidemics\\
Supplementary Material}
\author{Jewell CP \& Roberts GO}

\maketitle

%%%%%%%%%%%%% RELEVANT TEXT HERE %%%%%%%%%%%%

\section{Proof of Theorem 1}

Consider a time interval $[0,u)$ which is divided into
intervals $\Delta t=[t-\delta,t)$.   Let $\mathcal{W}_j = \{ \Delta t; 0 \leq t < u \}$ the set of time intervals for which contact tracing is observed, and $\mathcal{W}_j^c$ its complement.  Further we define the $\sigma-$algebras 
$\mathcal{F} = \sigma\left(\Delta C_{ijk}(t); t \in \mathcal{W}_{j}\right)$ and $\mathcal{G} = \sigma\left(\Delta C_{ijk}(t); t \in \mathcal{W}_{j}^c\right)$ 
representing the information contained in contact tracing data, and the information lost by not having contact tracing data, respectively.

If we were to observe contact tracing for all time (and not just a contact tracing window), a geometric likelihood may be used to make
inference on the vector of infectivity parameters $\bm{\xi}$, susceptibility parameters $\bm{\eta}$, and $p_k$ the probability that a contact along network $k=\{1,\dots,K\}$ of $K$ networks results in an infection.  Letting $\tilde{p}_{ijk}(t) = q(i;\bm{\xi},t)s(j;\bm{\eta})p_k$, $0 \leq \tilde{p}_{ijk}(t) \leq 1$, 

\begin{eqnarray}
L_\Omega(\bm{I},\bm{N},\bm{R}|\bm{\theta}) & = &  \prod_{j \in \mathcal{P}} \prod_{\Delta t \in [0,u)} \prod_{i \in \mathcal{Y}(t)}  \prod_{k=1}^{K} \left[ (\tilde{p}_{ijk}(t))^{\bm{1}[I_j \in \Delta t]} (1-\tilde{p}_{ijk}(t))^{1-\bm{1}[I_j < t]} \right]^{\Delta C_{ijk}(t)} \nonumber \\
 & \times &  \prod_{j:N\leq T_{obs}}f_{D}\left(N_{j}-I_{j}\right) \times \prod_{j:N_j>T_{obs}}\left[1-F_{D}\left(T_{obs}-I_{j}\right)\right] \nonumber \\
& = & \prod_{j \in \mathcal{P}}  \prod_{\Delta t \in [0,u)} \prod_{i \in \mathcal{Y}(t)} \prod_{k=1}^{K} e^{Z_{ijk}(t)\Delta C_{ijk}(t)} \nonumber \\
 & \times &  \prod_{j:N\leq T_{obs}}f_{D}\left(N_{j}-I_{j}\right) \times \prod_{j:N_j>T_{obs}}\left[1-F_{D}\left(T_{obs}-I_{j}\right)\right] \nonumber
\end{eqnarray}
where $\mathcal{Y}(t) = \mathcal{I}(t) \cup \mathcal{N}(t)$ the set of both \emph{infected} and \emph{notified} individuals at time $t$ and
\[ Z_{ijk}(t) = \left( \bm{1}[I_j \in \Delta t] \log(\tilde{p}_{ijk}(t)) + (1-\bm{1}[I_j \geq t+\delta])\log(1-\tilde{p}_{ijk}(t)) \right).  \]

Where contact tracing is not observed, for time intervals in $\mathcal{W}_j^c$, the presence or absence of a contact is replaced with the expectation with respect to $\mathcal{G}$
\begin{eqnarray}\label{eq:expectation}
L_{\mathcal{F}}(\bm{I},\bm{N},\bm{R}|\bm{\theta}) & = & \mathbb{E}_{\mathcal{G}} \left[  \prod_{j \in \mathcal{P}} \prod_{\Delta t \in \mathcal{W}_j} \prod_{i \in \mathcal{Y}(t)} \prod_{k=1}^{K} e^{Z_{ijk}(t)\Delta C_{ijk}(t)}
  \prod_{j \in \mathcal{P}} \prod_{\Delta t \in \mathcal{W}_j^c} \prod_{i \in \mathcal{Y}(t)} \prod_{k=1}^{K} e^{Z_{ijk}(t)\Delta C_{ijk}(t)} \right] \nonumber \\
& = & \prod_{j \in \mathcal{P}} \left[ \exp \left( \sum_{\Delta t \in \mathcal{W}_j} \sum_{i \in \mathcal{Y}(t)}\sum_{k=1}^{K} Z_{ijk}(t)\Delta C_{ijk}(t)\right) \right. \nonumber \\
& & \times \left. \mathbb{E}_{\mathcal{G}} \left[  \exp \left( \sum_{\Delta t \in \mathcal{W}_j^c} \sum_{i \in \mathcal{Y}(t)} \sum_{k=1}^{K} {Z_{ijk}(t)\Delta C_{ijk}(t)} \right) \right] \right]
\end{eqnarray}

We now address the question of how to calculate \[\mathbb{E}_{\mathcal{G}} \left[ \exp \left( \sum_{\Delta t \in \mathcal{W}_j^c} \sum_{i \in \mathcal{Y}(t)}\sum_{k=1}^{K} {Z_{ijk}(t)\Delta C_{ijk}(t)} \right) \right]\] for all $j$.  Conditional on the history $\mathcal{H}_t$ of the epidemic at time $t$, if we know the frequency $r_{ij}^{(k)}$ of contacts between individuals $i$ and $j$ along network $k$, we can write the joint probability of $j$ being infected via $k$ in the notional time interval $\Delta t=[t-\delta,t)$ 
\begin{eqnarray}
P\left(I_j \in \Delta t,\Delta C(t)=1|\mathcal{H}_{t}\right) & = & P\left(I_j \in \Delta t|\Delta C(t)=1,\mathcal{H}_{t}\right)P\left(\Delta C(t)=1|\mathcal{H}_{t}\right)\nonumber \\
 & = & \sum_{i\in \mathcal{I}(t)}\tilde{p}_{ijk}(t)r_{ij}^{(k)}\delta+o(\delta)\nonumber \\
\nonumber \\
\Rightarrow\mathbb{P}\left(I_j \not\in \Delta t|\mathcal{H}_{t}\right) & = & 1-\sum_{i\in\mathcal{I}(t)}\tilde{p}_{ijk}(t)r_{ij}^{(k)}\delta+o(\delta)\nonumber \end{eqnarray}

In terms of the probability density, this is equivalent to taking
an expectation over the contact tracing data $\mathcal{C}$ such that for all $j$
\begin{eqnarray*}
\mathbb{E}_{\mathcal{G}} \left[ \exp \left( \sum_{\Delta t \in \mathcal{W}_j^c} \sum_{i \in \mathcal{I}(t)} \sum_{k=1}^{K} {Z_{ijk}(t)\Delta C_{ijk}(t)} \right) \right] & = & \prod_{\Delta t \in \mathcal{W}_j^c} \left[ \left(\delta \lambda_j(t) \right)^{\bm{1}[I_j \in \Delta t]} e^{-\delta \lambda_j(t)} \right] \\
  &  & +o(\delta)
 \end{eqnarray*}
with $\lambda_{j}(t)=\sum_{k=1}^{K}\left[ \left(\sum_{i\in\mathcal{I}(t)}\tilde{p}_{ijk}(t)r_{ij}^{(k)} + \gamma \sum_{i\in\mathcal{N}(t)}\tilde{p}_{ijk}(t) r_{ij}^{(k)}\right)\right]$ the total infectious pressure on $j$ at time $t$. 

Notice that the first term in the square parentheses picks out intervals in which infections
have occurred (implying also that a contact occurred), with the remaining
intervals up to $j$'s infection time $I_j$ appearing in the second term. Observing that the number of intervals in which contacts
occurs becomes a negligible proportion of the total number of intervals,
the second terms becomes
\[
\lim_{\delta\rightarrow0} \prod_{\Delta t \in \mathcal{W}_j^c} \left[ \left(\delta \lambda_j(t) \right)^{\bm{1}[I_j \in \Delta t]} e^{-\delta \lambda_j(t)} \right]
 \propto \prod_{j: I_j < T_{obs}} \lambda_j(I_j^-) \exp \left( -\sum_{j \in \mathcal{P}} \int_{I_\kappa}^{I_j \wedge T_{obs}} \lambda_j(t)dt \right)
\]
This is equivalent to the form of the exponential likelihood in Equation 1, main paper.  

Taking the limit $\delta \rightarrow 0$ throughout and a product over all individuals in the population -- since in the Markovian structure of the epidemic model infection events are conditionally independent -- the embedded contact tracing likelihood is

\begin{eqnarray}
L_{\mathcal{F}}(\bm{I},\bm{N},\bm{R} | \bm{\theta}) & = & \prod_{j:I_j \in \mathcal{W}_j} \exp \left[ \int_{\mathcal{W}_j} \sum_{i\in \mathcal{Y}(t)} \sum_{k=1}^{K} Z_{ijk}(t) dC_{ijk}(t) \right]\nonumber \\
& \times & \prod_{\substack{j:I_j\in \mathcal{W}_j^c \\ j\ne\kappa}} \sum_{i\in\mathcal{Y}(I_{j}^{-})}\lambda_j(I_j^-)\cdot\exp\left[-\int_{\mathcal{W}_j^c}\lambda_j(t)dt\right]
\end{eqnarray}

\paragraph{Remarks} For networks, or more generally modes of disease transmission, where contact frequency is not known $p_k$ and $r_{ij}^{(k)}$ cannot be identified.  Instead, we make inference on the overall infection rate $\beta_k = p_k r_{ij}^{(k)}$, where it is assumed that $r_{ij}^{(k)}$ is constant with respect to both $i$ and $j$.  These modes of transmission necessarily appear in the $\mathcal{W}_j^c$ terms in the likelihood, and are therefore unable to admit contact tracing data in our current setup.

%%%%%%%%%%%%%% RELEVANT BIT ENDS HERE %%%%%%%%%%%%%%

\pagebreak

\section{MCMC Algorithm\label{app:MCMC-Algorithm}}

The following reversible jump MCMC algorithm is used to estimate the
joint posterior distribution of the Bayesian epidemic model described
in Sections 4 and 5 of the main paper. It is based on the methodology developed in
\citet{JewEtAl09c}, but contains significant methodological enhancements
to encompass the binomial addition to the likelihood. Specifically,
since infections can now be caused by either an observed
or unobserved contact, MCMC updates for the infection
times need to be modified, taking into account that for any particular
iteration, an infection time can exist in one of two states: $I_j \in \mathcal{W}_j$ if $j$ is infected by an observed contact, 
or $I_j \in \mathcal{W}_j^c$ if $j$ is infected in a period of time for which contact tracing data is not observed.

Defining the number of occult infections at iteration $n$ as $m^{(n)} = \sum_{i \in \bm{I^{(n)}}} \ind(N_i > T_{obs})$, the algorithm proceeds as follows:
\begin{enumerate}
\item Take starting values $\bm{\theta}^{(0)} =\left\{ \epsilon^{(0)}, \bm{p}^{(0)}, \bm{\beta}^{(0)}, \gamma^{(0)},\psi^{(0)}, \bm{\eta}^{(0)} \right\}$ and $\bm{I}^{(0)}$
\item Update $\bm{\theta}^{(n+1)}|\bm{I}^{(n)},\bm{N}^{(n)},\bm{R}^{(n)},\mathcal{C}^{(n)}$
\item For $z$ times choose one of:

\begin{enumerate}
\item Update an infection time: $I_{s}^{(n+1)}|\bm{I}_{-s}^{(n)},\bm{N}^{(n)},\bm{R}^{(n)}$ with
$s\sim\mbox{Uniform}([\bm{I}^{(n)}])$
\item Add an infection: $\bm{I}^{(n+1)} = \{\bm{I}^{(n)}+s\}$ with $s\sim\mbox{Uniform}([\mathcal{S}^{(n)}])$ 
\item Delete a previously added infection: $\bm{I}^{(n+1)} = \{\bm{I}^{(n)}-s\}$ with $s\sim\mbox{Uniform}([\mathcal{I}^{(n)}(T_{obs})])$
\end{enumerate}
\item Goto 2
\end{enumerate}
In step (2), the entire transmission parameter vector $\bm{\theta}$
 is updated using an adaptive multisite Metropolis-Hastings
step \citep{HaarSakTam01}. The steps to
move, add, or delete an infection are now described in detail.

\subsection{Moving an infection time}

The derivation of the contact tracing likelihood implies
that if $I_{s}$ is equal to the time of a known contact arriving
at $s$, then the contact must have
caused the infection; conversely, if $I_{s}$ falls at any other time, then it must have been infected
by a rate-driven component of the model. Since $I_{s}$ appears in
different parts of the likelihood depending on $\mathcal{W}_j$,
four possibilities exists for moving an infection time:
\begin{lyxlist}{00.00.0000}
\item [{(a)}] \textbf{$I_{s}\in\mathcal{W}_s\rightarrow I_{s}^{\star}\in\mathcal{W}_s$:
A contact infection time is moved to another contact infection time. }
\item [{(b)}] \textbf{$I_{s}\in\mathcal{W}_s^c\rightarrow I_{s}^{\star}\in\mathcal{W}_s^c$:
Non-contact infection time is moved to another non-contact infection
time.}
\item [{(c)}] \textbf{$I_{s}\in\mathcal{W}_s^c\rightarrow I_{s}^{\star}\in\mathcal{W}_s$:
Non-contact infection time is moved to a contact infection time.}
\item [{(d)}] \textbf{$I_{s}\in\mathcal{W}_s\rightarrow I_{s}^{\star}\in\mathcal{W}_s^c$:
Contact infection time is moved to a non-contact infection time.}
\end{lyxlist}

Moves (c) and (d) result in $I_{s}$ appearing in different parts
of the likelihood, resulting in a change of model dimension. These
moves will therefore be termed ``\emph{trans-dimensional}''. Conversely,
moves (a) and (b) do not change the dimension of the model and are
therefore termed ``\emph{cis-dimensional}''. Given the current
status of $I_{s}$, a \emph{trans-dimensional }move is proposed with
probability 0.5, else a \emph{cis-dimensional} move occurs. For proposing
a contact infection time, the following distribution is used
\[
I_{s}^{\star}\sim\mbox{Uniform}\left(\mathcal{\tilde{C}}_s^{(n)}\right)
\]
where $\mathcal{\tilde{C}}_s^{(n)}$ is
the set of $s$'s incoming contacts (on all networks) that originate from \emph{infective}
individuals $i$ at the current state $n$ of the Markov chain (ie
known infectives and occults).  In other words, we propose a new infection time by sampling uniformly from the contact times.  
Note that the
only valid move for $s:\mathcal{W}_s = \emptyset$, which includes
occults, is (b).  

In conjunction with the four possible proposal schemes, therefore,
there are four possible acceptance probabilities for updating an infection time chosen uniformly at random:
\begin{lyxlist}{00.00.0000}
\item [{(a)}] $I_{s}\in\mathcal{W}_s\rightarrow I_{s}^{\star}\in\mathcal{W}_s$
: Propose $I_{s}^{\star}\sim\mbox{Uniform}\left(\mathcal{\tilde{C}}_s^{(n)}\right)$
and accept with probability\\
\begin{eqnarray*}
\alpha(I_{s}^{(n)},I_{s}^{(n+1)}) & = & 1\wedge\left[\frac{\pi(\bm{I}^{\star}|\bm{N},\bm{R},\bm{\mathcal{C}}^{(n)},\bm{\theta}^{(n)})}{\pi(\bm{I}^{(n)}|\bm{N},\bm{R},\bm{\mathcal{C}}^{(n)},\bm{\theta}^{(n)})}\times\frac{\left[\mathcal{\tilde{C}}_s^{(n)}\right]}{\left[\mathcal{\tilde{C}}_s^{(n+1)}\right]}\right]\\
 & = & 1\wedge\frac{\pi(\bm{I}^{\star}|\bm{N},\bm{R},\bm{\mathcal{C}}^{(n)},\bm{\theta}^{(n+1)})}{\pi(\bm{I}^{(n)}|\bm{N},\bm{R},\bm{\mathcal{C}}^{(n)},\bm{\theta}^{(n+1)})}\end{eqnarray*}

\item [{(b)}] $I_{s}\in\mathcal{W}_s^c\rightarrow I_{s}^{\star}\in\mathcal{W}_s^c$:
Propose $I_{s}^{\star}\overset{D}{=}g_{D}(N_{s}-I_{s}^{\star})$ if
known and $I_{s}^{\star}\overset{D}{=}\tilde{g}_{D}(T_{obs}-I_{s}^{\star})$
if occult. Accept with probability\\
\[
\alpha(I_{s}^{(n)},I_{s}^{(n+1)})=\begin{cases}
1\wedge\left[\frac{\pi(\bm{I}^{(n+1)}|\bm{N},\bm{R},\mathcal{C}^{(n)},\bm{\theta}^{(n)})}{\pi(\bm{I}^{(n)}|\bm{N},\bm{R},\mathcal{C}^{(n)},\bm{\theta}^{(n)})}\right] \times \frac{g_D(N_s - I_s^{(n)})}{g_D(N_s - I_s^\star)} & \mbox{if }j\mbox{ is known}\\
1\wedge\left[\frac{\pi(\bm{I}^{(n+1)}|\bm{N},\bm{R},\mathcal{C}^{(n)},\bm{\theta}^{(n)})}{\pi(\bm{I}^{(n)}|\bm{N},\bm{R},\mathcal{C}^{(n)},\bm{\theta}^{(n)})}\right] \times \frac{\tilde{g}_D(T_{obs} - I_s^{(n)})}{\tilde{g}_D(T_{obs} - I_s^\star)} & \mbox{if }j\mbox{ is occult}\end{cases}\]

\item [{(c)}] $I_{s}\in\mathcal{W}_s^c\rightarrow I_{s}^{\star}\in\mathcal{W}_s$:
Propose $I_{s}^{\star}\sim\mbox{Uniform}\left(\mathcal{\tilde{C}}_s^{(n)}\right)$
and accept with probability\\
\[
\alpha(I_{s}^{(n)},I_{s}^{(n+1)})=1\wedge\left[\frac{\pi(\bm{I}^{(n+1)}|\bm{N},\bm{R},\mathcal{C}^{(n)},\bm{\theta}^{(n)})}{\pi(\bm{I}^{(n)}|\bm{N},\bm{R},\mathcal{C}^{(n)},\bm{\theta}^{(n)})}\times\frac{g_{D}(N_{s}-I_{s}^{(n)})}{1/\left[\mathcal{\tilde{C}}_s^{(n)}\right]}\right]\]

\item [{(d)}] $I_{s}\in\mathcal{W}_s\rightarrow I_{s}^{\star}\in\mathcal{W}_s^c$:
Propose $I_{s}^{\star}\overset{D}{=}g_{D}(N_{s}-I_{s}^{\star})$\\
\[
\alpha(I_{s}^{(n)},I_{s}^{(n+1)})=1\wedge\left[\frac{\pi(\bm{I}^{(n+1)}|\bm{N},\bm{R},\mathcal{C}^{(n)},\bm{\theta}^{(n)})}{\pi(\bm{I}^{(n)}|\bm{N},\bm{R},\mathcal{C}^{(n)},\bm{\theta}^{(n)})}\times\frac{1/\left[\mathcal{\tilde{C}}_s^{(n)}\right]}{g_{D}(N_{s}-I_{s}^{\star})}\right]\]

\end{lyxlist}

Probability densities $g_D(\cdot)$ and $\tilde{g}_D(\cdot)$ are used for proposing non-contact infection times for known and occult infections respectively.  Similarly to our previous work \citep{JewEtAl09c}, we propose from $g_D(t) \overset{D}{=} f_D(t)$ for known infections, and approximate the censored density $1-F_D(t)$ using a second order Taylor expansion
\[
\tilde{g}_D(t) \overset{D}{=} \mbox{ Normal}\left( -\frac{1}{b}, \frac{1}{ab^2} \right)
\]

\subsection{Adding or deleting an infection time}

The addition and deletion updates are performed as in \citet{JewEtAl09c}.
Briefly, for an addition, an individual $s$ is chosen uniformly from
$\mathcal{S}^{(n)}$, the set of susceptible individuals at the $n$th
iteration of the Markov chain. An infection time is then proposed
from the density $1-F_{D}(T_{obs}-I_{s}^{\star})$ as for moving an
occult infection time. The addition is accepted with probability\[
\alpha\left(\bm{I}^{(n)},\bm{I}^{(n+1)}\right)=1\wedge\frac{\pi\left(\left\{ \bm{I}^{(n)}+s\right\} |{\bm{N}},{\bm{R}},\bm{\beta}^{(n)}\right)}{\pi(\bm{I}^{(n)}|{\bm{N}},{\bm{R}},\bm{\beta}^{(n)})}\times\frac{\left[\mathcal{S}^{(n)}\right]}{\left(\left[\mathcal{S}^{(n)}\right]+1\right)\cdot\tilde{g}_{D}(T_{obs}-I_{s}^{\star})}\]

For a deletion, an occult individual $s$ is chosen uniformly. It's
deleting is then accepted with probability\[
\alpha\left(\bm{I}^{(n)},\bm{I}^{(n+1)}\right)=1\wedge\frac{\pi\left(\left\{ \bm{I}^{(n)}-s\right\} |{\bm{N}},{\bm{R}},\bm{\beta}^{(n)}\right)}{\pi(\bm{I}^{(n)}|{\bm{N}},{\bm{R}},\bm{\beta}^{(n)})}\times\frac{\left[\mathcal{S}^{(n)}\right]\cdot\tilde{g}_{D}(T_{obs}-I_{s})}{\left[\mathcal{S}^{(n)}\right]-1}\]

The sets $\mathcal{\tilde{C}}_{s}^{(n)}$ are consequentially updated for all $j$ since contacts
that were previously thought to have originated at susceptibles, might
now originate at infecteds, and vice versa.

\pagebreak
\section{Supporting Figures}

\begin{figure}[H]
\noindent \begin{centering}
\includegraphics[width=0.8\textwidth]{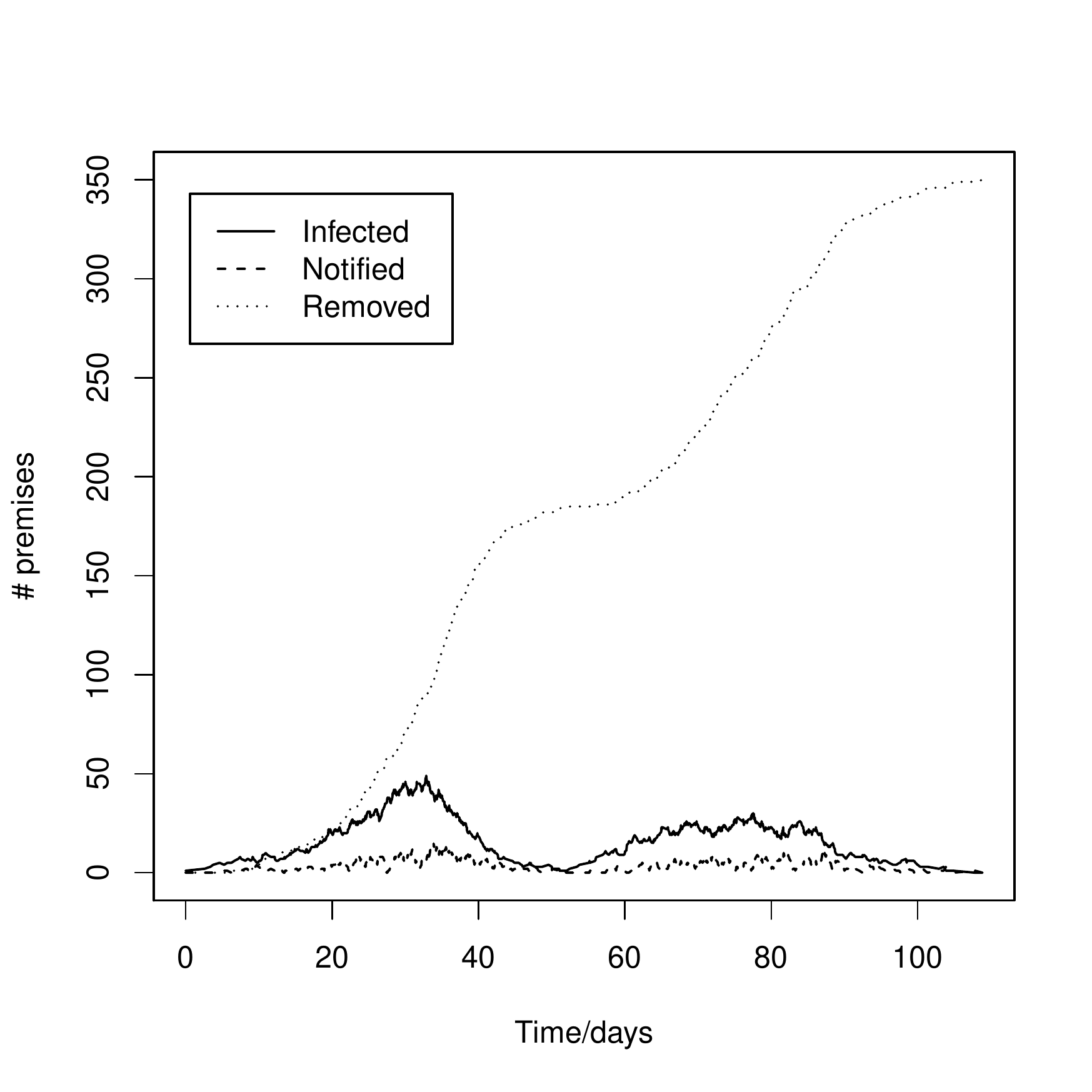}
\caption{\label{fig:parmEpi}Simulated epidemic trajectory for Section 6 of the main paper showing Infected,
Notified, and Removed individuals.}
\end{centering}
\end{figure}

\begin{figure}[H]
\noindent \begin{centering}
\includegraphics[width=\textwidth]{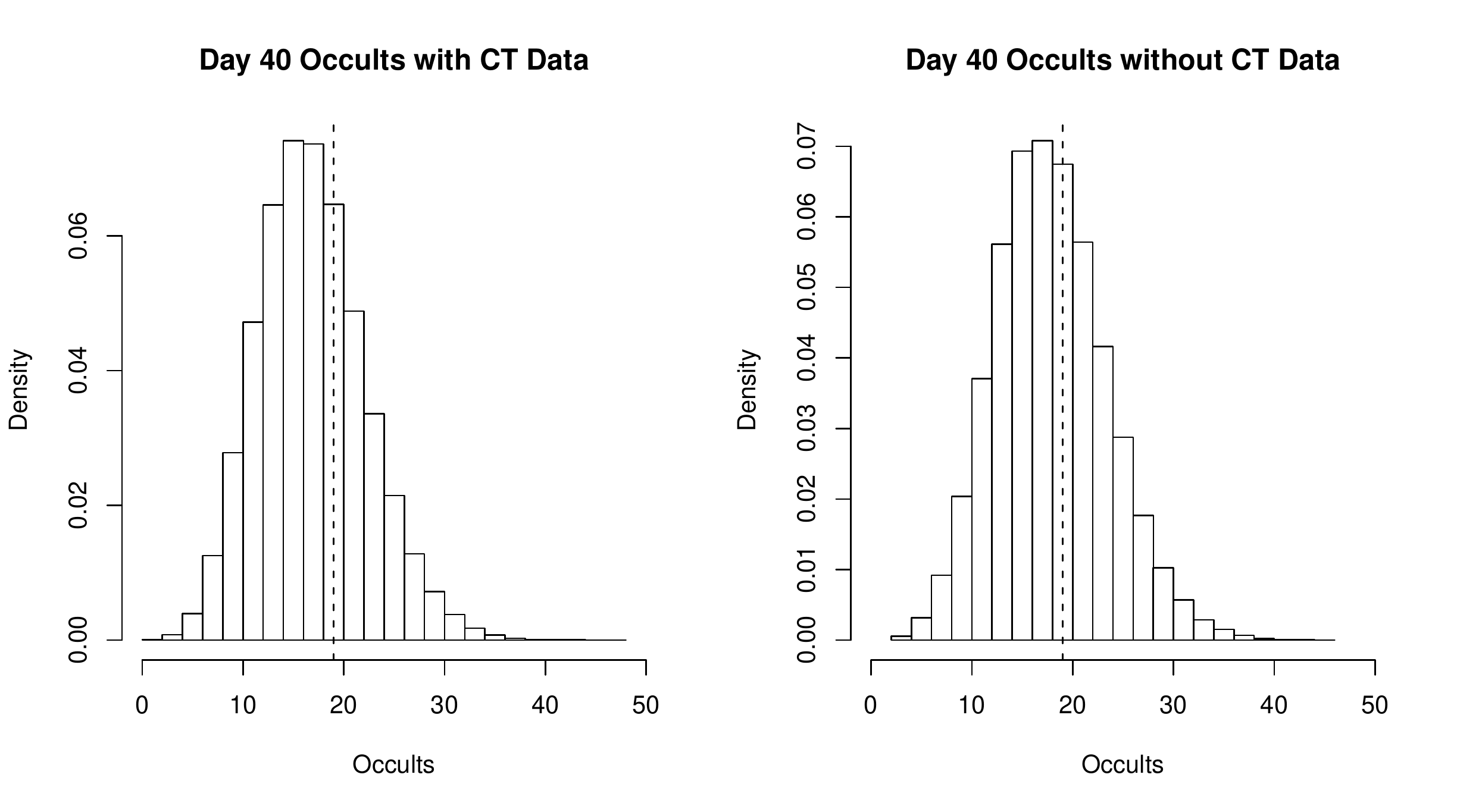}
\par\end{centering}

\caption{\label{fig:occultHist}Histogram of number of occult infections for
the test epidemic on day 40: with contact tracing (left); without
contact tracing (right). True number of occult infections indicated
by dashed line.}

\end{figure}

\begin{figure}[H]
\noindent \begin{centering}
\includegraphics[width=\textwidth]{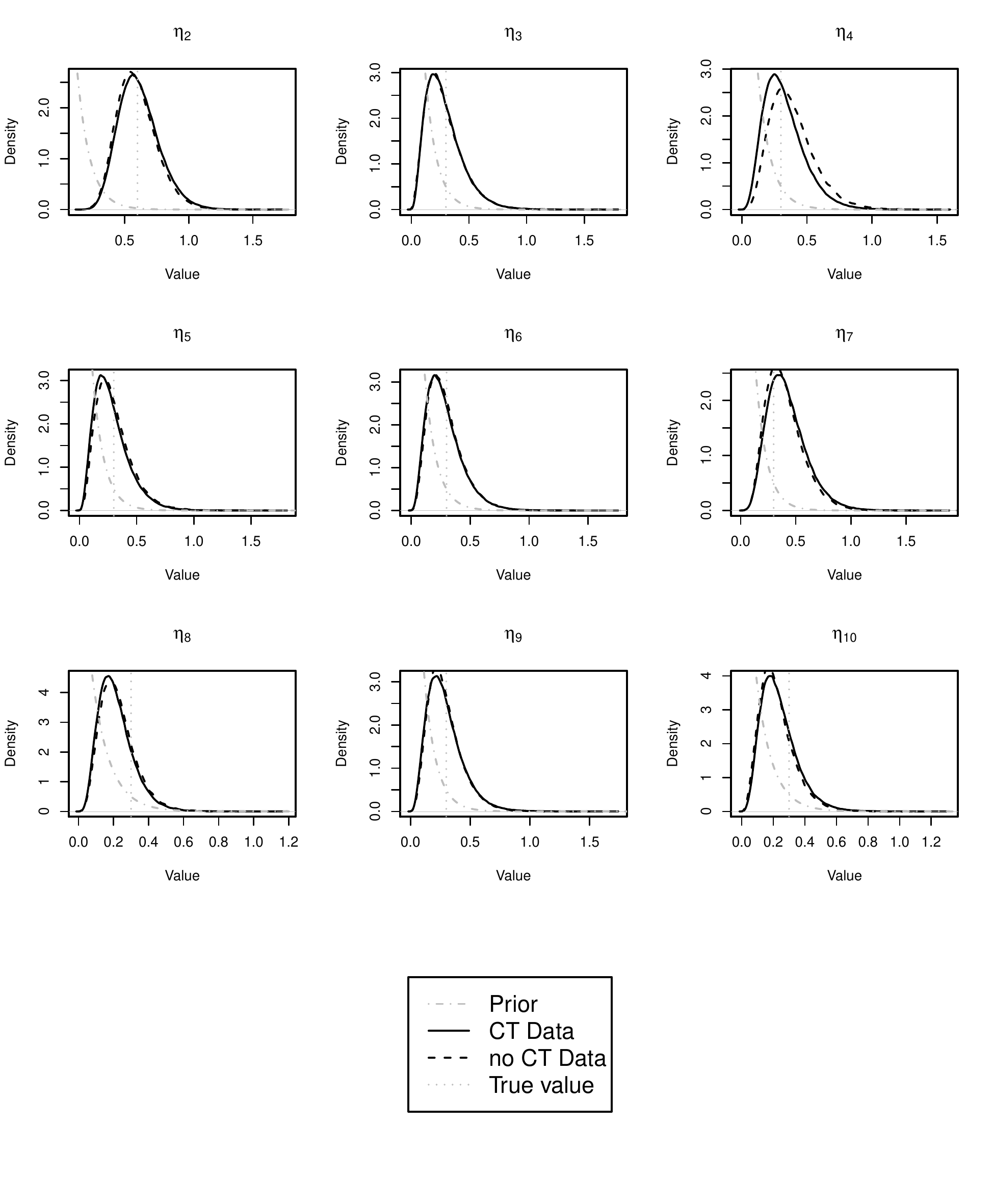}
\par\end{centering}

\caption{Production type susceptibilities ($\eta_{2\dots9}$) relative to broiler
chickens ($\eta_{1}$) for the Day 40 example.}

\end{figure}

\begin{figure}[H]
\noindent \begin{centering}
\includegraphics[width=\textwidth]{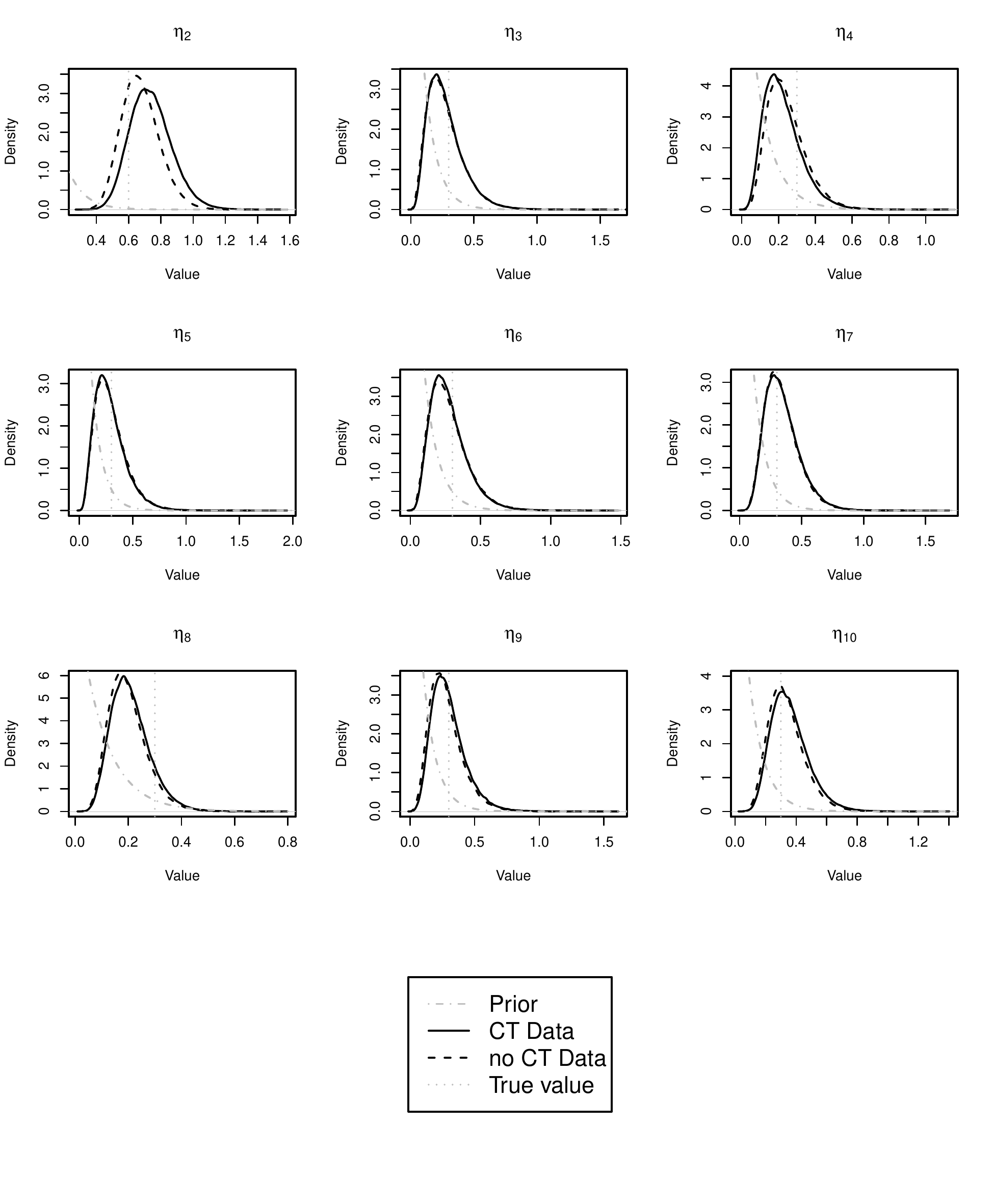}
\par\end{centering}

\caption{Production type susceptibilities ($\eta_{2\dots9}$) relative to broiler
chickens ($\eta_{1}$) for the full epidemic (Day 109) example.}

\end{figure}

\begin{figure}[H]
\noindent \begin{centering}
\includegraphics[width=\textwidth]{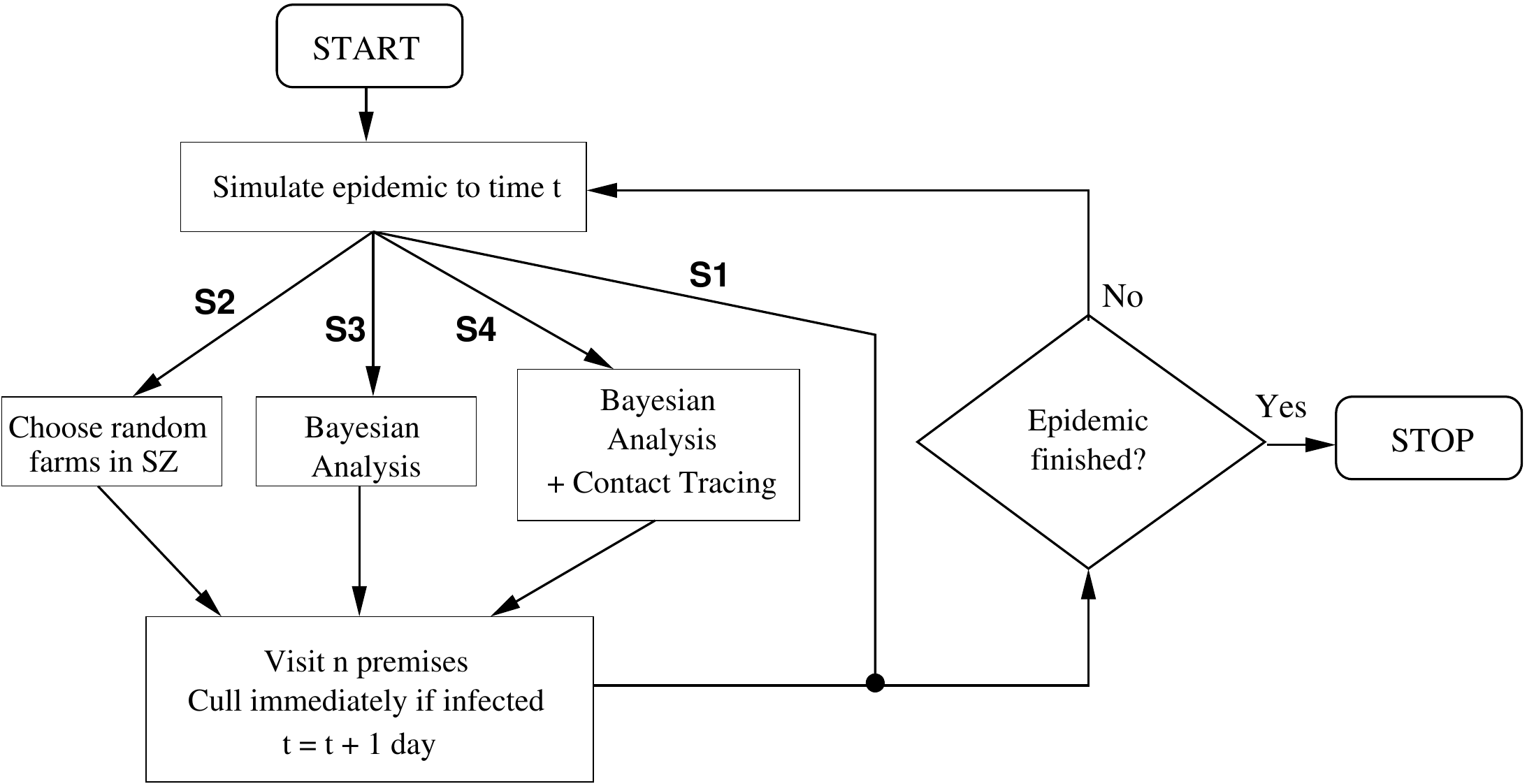}
\par\end{centering}

\caption{\label{fig:sim-dataflow}Dataflow diagram of surveillance strategy
simulation study. S1) basic {}``SOS' disease control; S2) random
surveillance within SZ; S3) Bayesian targeted surveillance, not using
contact tracing; S4) Bayesian targeted surveillance using contact
tracing data. }

\end{figure}

\begin{figure}[H]
\centering
\includegraphics[width=0.5\textwidth]{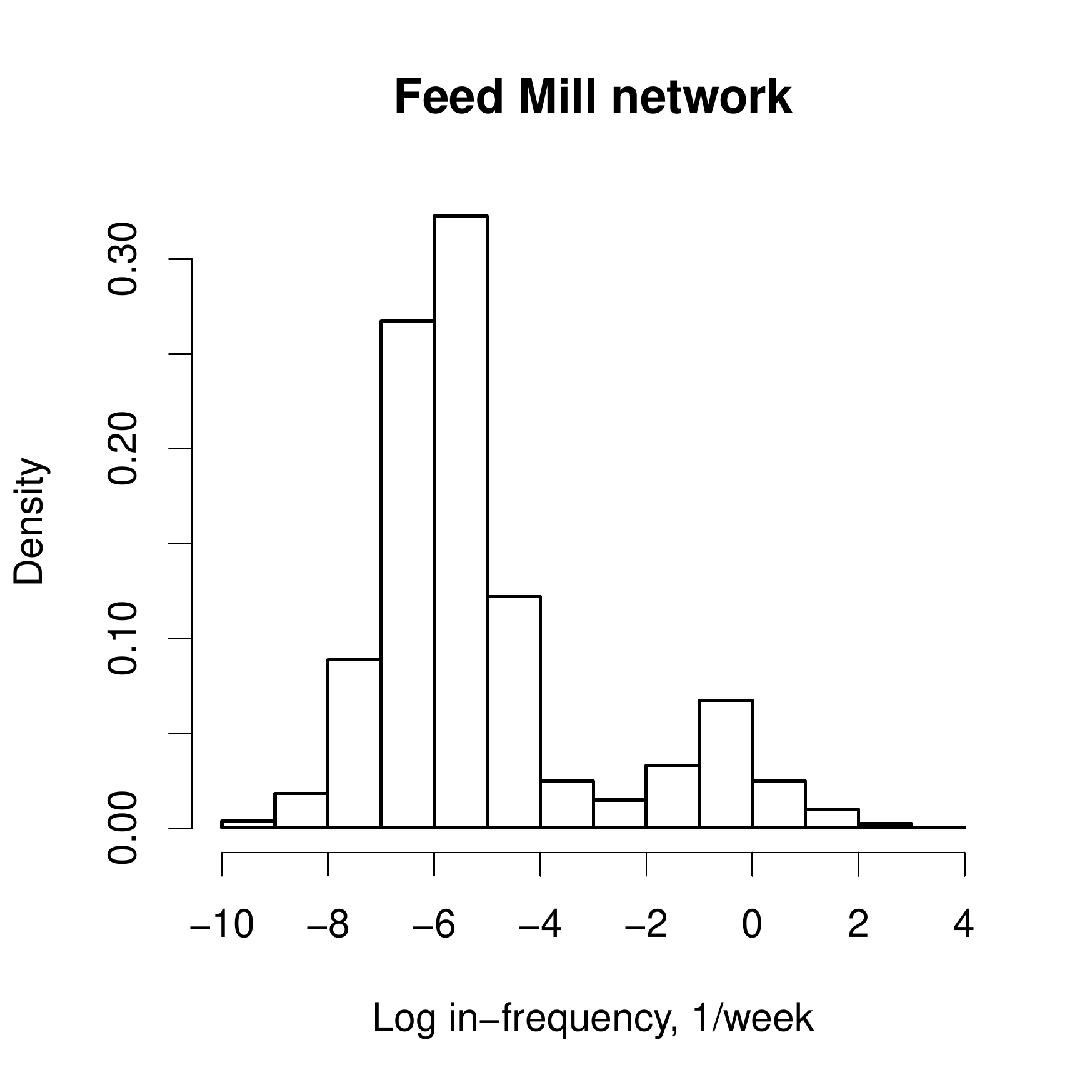} \\ \includegraphics[width=0.5\textwidth]{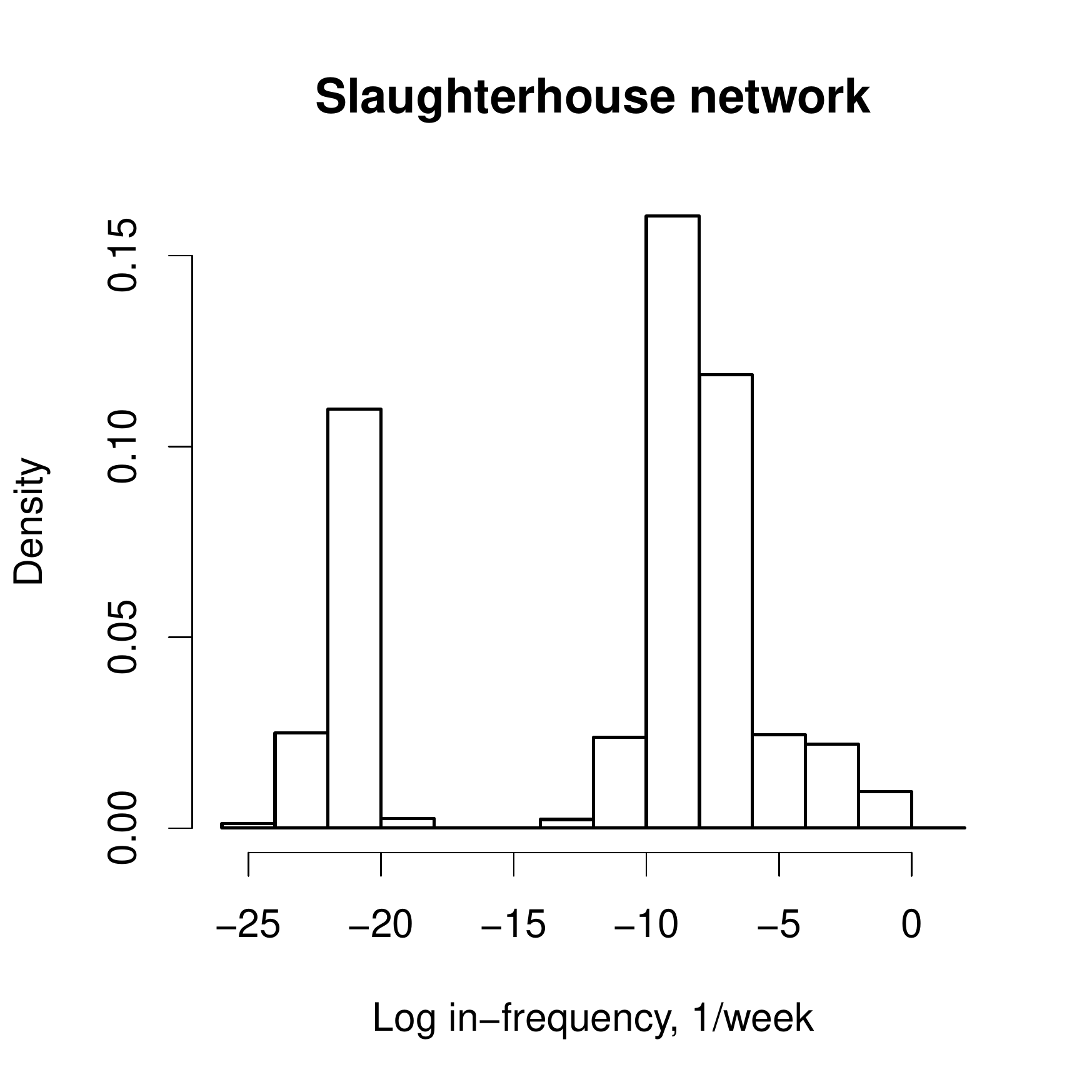}
\caption{\label{fig:netFreqs}Histograms of the log distribution of incoming contact frequency (contacts per week) for the feedmill (top) and slaughterhouse (bottom) networks, according to \citet{ShEtAl08}.}
\end{figure}

\section{Information gained from contact tracing data versus case incidence data}

To provide an intuition for the difference in added posterior information
between $p_{1}$ and $p_{2}$, it is useful to consider how the increase
in precision due to CTD varies as a function of the
underlying contact rate. In a simple model in which an individual
gets infected via a frequency network with contact rate $r$ and infection probability $p$ 
during an interval of length $\delta$, the variance of the estimator $\hat{p}$ is
 $$\mbox{Var}_{exp}(\hat{p})=1/(r\delta)^2$$ for an exponential model and 
$$\mbox{Var}_{geo}(\hat{p})=1/(r\delta)^2 - 1/(r\delta)^3$$ for the geometric model.  $\mbox{Var}_{geo}(\hat{p})$ is therefore guaranteed less than $\mbox{Var}_{exp}(\hat{p})$.

This gives insight into the fact that the acquisition of any contact
tracing data always adds statistical information, and since
for large $r$ the first term in $\mbox{Var}_{geo}(\hat{p})$ 
dominates, the variance converges for the two models as shown below in Figure \ref{fig:ctVar}. In other words, although the variance
of the geometric model is always less than for the exponential model,
the inclusion of CTD may only be worthwhile if contact
rates are low.

\begin{figure}[H]
\centering
\includegraphics[width=0.8\textwidth]{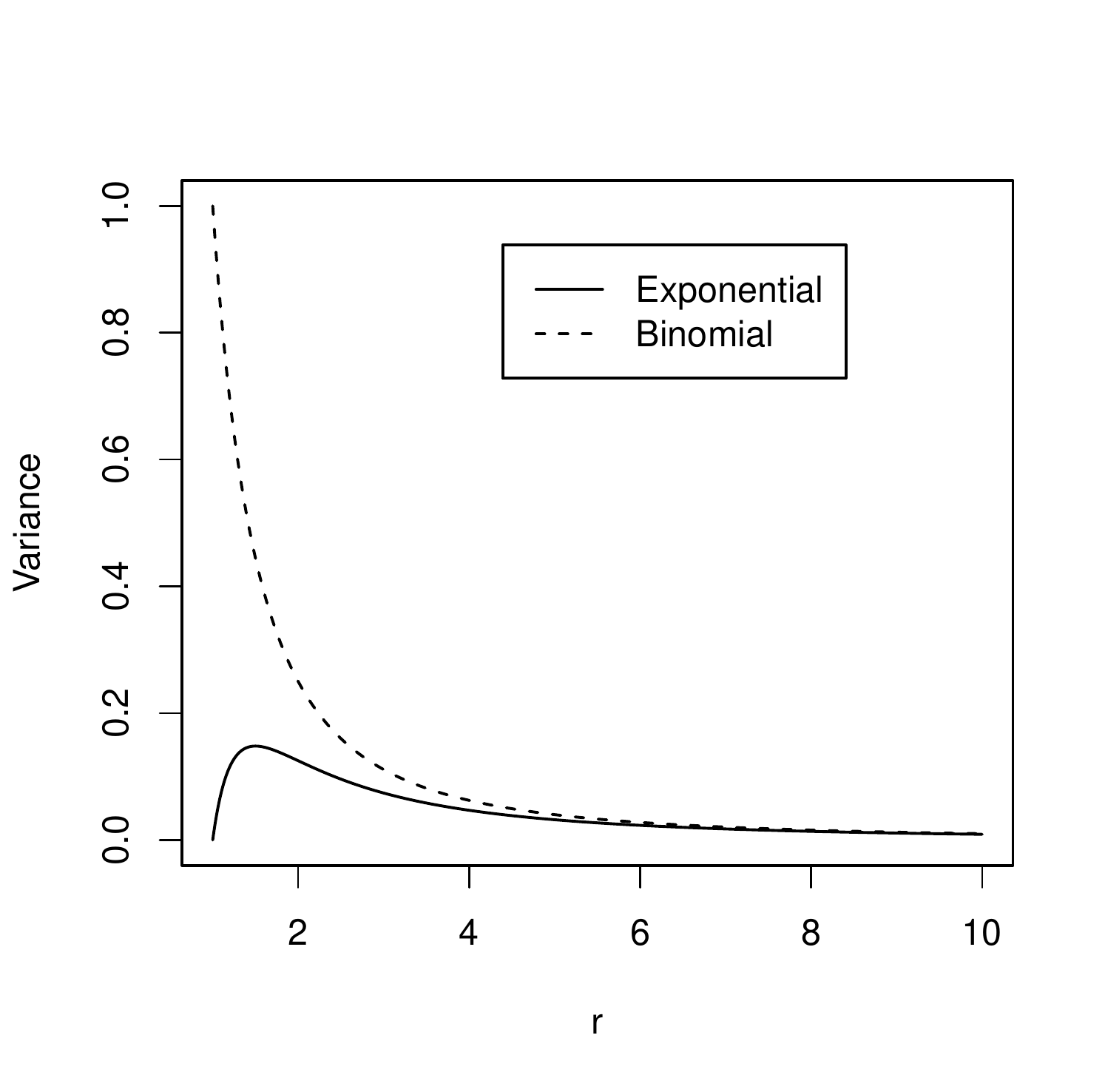}
\caption{\label{fig:ctVar}Var($p$) for both geometric and exponential models with $\delta=1$.}
\end{figure}

\bibliographystyle{apalike}
\bibliography{general}